\documentclass[11pt]{article}
\usepackage{graphicx}
\renewcommand{\baselinestretch}{1.4}
\begin{document}
\title{\bf  The limits of the total crystal-field splittings}
\author{\bf J. Mulak$^{1}$, M. Mulak$^{2}$}
\date{{\it  $^{1}$ Trzebiatowski Institute of Low Temperature
            and Structure Research,\\
            Polish Academy of Sciences, 50--950, PO Box 1410,
            Wroclaw, Poland\\
            $^{2}$ Institute of Physics,
            Wroclaw University of Technology,\\
            Wyb. Wyspianskiego 27,
            50--370 Wroclaw, Poland}}
\maketitle
\vspace*{0.2cm}
\noindent
{\bf Corresponding author:}\\
Prof. Jacek Mulak\\
Trzebiatowski Institute of Low Temperature and
Structure Research,\\
Polish Academy of Sciences,
50--950, PO Box 1410,
Wroclaw, POLAND\\
email: Maciej.Mulak@pwr.wroc.pl\\
Tel: (+4871) 3435021, 3443206\\
Fax: (+4871) 3441029\\
\begin{abstract}
\noindent The crystal-fields causing $|J\rangle$ electron states splittings of the same second moment
$\sigma^{2}$ can produce different total splittings $\Delta E$ magnitudes. Based on the numerical data on
crystal-field splittings  for the representative sets of crystal-field Hamiltonians ${\cal H}_{\rm
CF}=\sum_{k}\sum_{q}B_{kq}C_{q}^{(k)}$ with fixed indexes either $k$ or $q$, the potentials leading to the
extreme $\Delta E$ have been identified. For all crystal-fields the admissible ranges $(\Delta E_{min},\Delta
E_{max})$ have been found numerically for  $1\leq J\leq 8$. The extreme splittings are reached in the
crystal-fields for which ${\cal H}_{\rm CF}s$ are the definite superpositions of the $C_{q}^{(k)}$ components
with different rank $k=2,4$ and $6$ and the same index $q$. Apart from few exceptions, the lower limits $\Delta
E_{min}$ occur in the axial fields of ${\cal H}_{\rm
CF}(q=0)=B_{20}C_{0}^{(2)}+B_{40}C_{0}^{(4)}+B_{60}C_{0}^{(6)}$, whereas the upper limits $\Delta E_{max}$ in
the low symmetry fields of ${\cal H}_{\rm CF}(q=1)=B_{21}C_{1}^{(2)}+B_{41}C_{1}^{(4)}+B_{61}C_{1}^{(6)}$.
Mixing the ${\cal H}_{\rm CF}$ components with different $q$ yields a secondary effect and does not determine
the extreme splittings. The admissible $\Delta E_{min}$ changes with $J$ from $2.00\sigma$ to $2.40\sigma$,
whereas the $\Delta E_{max}$ from $2.00\sigma$ to $4.10\sigma$. The maximal gap $\Delta E_{max}-\Delta
E_{min}=2.00\sigma$ has been found for the states $|J=4\rangle$. Not all the nominally allowed total splittings,
preserving $\sigma^{2}=const$ condition, are physically available, and in consequence not all virtual splittings
diagrams can be observed in real crystal-fields.
\end{abstract}

\section*{1. Introduction}
The total splitting ranges $(\Delta E_{min},\Delta E_{max})$ of free ion electron states in different axial
crystal-fields (CFs) yielding the splittings of the same second moment $\sigma^{2}$ have been thoroughly
analyzed in our previous paper [1]. Here, we extend this analysis for all CFs. Throughout the paper the tensor
Wybourne notation [2] for the CF Hamiltonian  ${\cal H}_{\rm CF}$ and the CF parameters $B_{kq}$ (CFP) is
consistently used (Eq.(1)),
\begin{equation}
{\cal H}_{\rm CF}=\sum_{k}\sum_{q}B_{kq}C_{q}^{(k)}.
\end{equation}
The zero-approximation initial states forming the interaction matrices are the Russell-Saunders coupled states
$|\alpha SLJ\rangle$ of well defined quantum number $J$ with the degeneracy $2J+1$, where $\alpha$ stands for
the remaining needed quantum numbers. For the sake of simplicity the initial states are denoted as $|J\rangle$.
Further extension of the studies for the mixed states as well as the states resulting from other coupling
schemes are feasible taking into account the additivity of the CF effect with respect to $2^{k}$-pole ${\cal
H}_{\rm CF}$ components. To achieve this one should employ consequently the tensor transformational properties
of the states along with the standard angular momentum re-coupling techniques [3-5]. The systematic calculations
have been performed in the paper for all the non-Kramers and Kramers states with $1\leq J\leq 8$, it means for
fifteen $J$ values.

In order to compare the splitting effects of arbitrary (real or virtual) ${\cal H}_{\rm CF}$s they must yield
the splitting diagrams of the same second moment $\sigma^{2}$ defined as [6-9]
\begin{equation}
\sigma^{2}|J\rangle=\frac{1}{2J+1}\sum\limits_{n} \left[ E_{n}-\overline{E}\left(|
J\rangle\right)\right]^{2}=\frac{1}{2J+1}\sum\limits_{k}S_{k}^{2} \left(\langle J||C^{(k)}||
J\rangle\right)^{2},
\end{equation}
where the energy centre of gravity of the sublevels within the state $|J\rangle$ is given by
$\overline{E}\left(| J\rangle\right)=\frac{1}{2J+1}\sum_{n}E_{n}$, with $E_{n}$ as the energy of $|n\rangle$
sublevel, and $S_{k}=\left(\frac{1}{2k+1}\sum_{q}|B_{kq}|^{2}\right)^{1/2}$ is the conventional CF strength of
the $2^{k}$-pole [6,7,10,11], whereas the dimensionless scalar $\langle J||C^{(k)}||J\rangle$ describes the
$2^{k}$-pole type aspherity of the state $|J\rangle$ [2,9].

To put in order different isomodular $2^{k}$-pole ${\cal H}_{\rm CF}^{(k)}$ components of the global ${\cal
H}_{\rm CF}$ the complementary scale of the CF strength was applied. This is the spherically averaged modulus of
${\cal H}_{\rm CF}^{(k)}$ which is equal to the spherically averaged modulus of the axial $B_{k0}$ CFP [8]
\begin{equation}
|{\cal H}_{\rm CF}^{(k)}|_{\rm \;av}=\frac{1}{4\pi}\int\limits_{0}^{2\pi}\int\limits_{0}^{\pi} |{\cal H}_{\rm
CF}^{(k)}(\alpha, \beta)|\;\sin\beta \;d\beta \;d\alpha=|B_{k0}|_{\rm \;av},
\end{equation}
where $\alpha$ and $\beta$ are two Euler angles of the reference frame rotation (the third angle is
inessential).

The most general ${\cal H}_{\rm CF}$ contains 27 CFPs including 3 real axial parameters and 12 pairs of complex
ones [2,12,13]. An interesting question arises whether one should consider complex parametrized ${\cal H}_{\rm
CF}$s investigating the extreme splittings $\Delta E_{min}$ and $\Delta E_{max}$ of electron states in any CF
yielding the same $\sigma^{2}$. The negative answer can be explained as follows. The off-diagonal matrix
elements of the complex ${\cal H}_{\rm CF}s$ are superpositions of at most three complex terms (for $k=2,4,6$),
which due to their own different phases can be treated as vector sums. On the other hand in the case of the real
parametrizations we deal solely with algebraic sums of the components. In consequence, the complex
parametrizations (with the phase angles as their additional degrees of freedom) considerably enlarge the set of
available ratios between the matrix elements comparing to a more discrete set for the real parametrizations. For
each complex parametrized ${\cal H}_{\rm CF}$ an equivalent real CF interaction matrix, leading to the same
splitting diagram, can be given. The point is, however, that such a matrix cannot be attained only by means of
the real CFPs. The complex Hamiltonians are therefore more capable than the real ones. This is what a supremacy
of the complex parametrizations over the real ones consists in. Since the absolute values of the vector sums are
enclosed within the range of the absolute values of the relevant algebraic sums, ${\cal H}_{\rm CF}s$ of complex
parametrizations are not relevant to $\Delta E_{min}$ and $\Delta E_{max}$, although they essentially enrich the
set of available splitting diagrams. Thus, searching for the extreme splittings we can confine ourselves
exclusively to the ${\cal H}_{\rm CF}$ of real CFPs. For the sake of simplicity the terms
$\frac{1}{\sqrt{2}}\left(C_{q}^{(k)}+C_{-q}^{(k)}\right)$ are written as $C_{q}^{(k)}$.

Clearly, the numerically-analytical approach applied previously to the three-parameter axial CFs, it this case
seems to be hardly realizable considering the large number of independent CFPs. In fact, for fifteen real CFPs
and fixed $\sigma^{2}$ one would need to scan independently and simultaneously all the involved $B_{kq}s$ within
the ranges from $0$ to $\sqrt{(2J+1)(2k+1)}$ (in $\sigma$), and then, after diagonalization of the ${\cal
H}_{\rm CF}$ interaction matrix, to find out the upper and lower limits of the dominating absolute differences
between the sublevels energies. In geometrical interpretation it would correspond to the analysis of adequate
number of plane equations (equal to the number of the sublevels) within the 15-dimensional space with respect to
the sphere of the radius $R=\sigma$ [1]. In order to avoid such difficulties we have investigated $\Delta
E_{min}$ and $\Delta E_{max}$ for a general ${\cal H}_{\rm CF}$ (Eq.(1)) employing an intermediate method.
Namely, we utilized a seemingly trivial fact that the CF effect is essentially differentiated for various
combinations of $q$ and $k$ indexes. Firstly, we have found the limits of the total splittings which can be
obtained separately for isomodular partial ${\cal H}_{\rm CF}^{(k)}=\sum_{q}B_{kq}C_{q}^{(k)}$ for $k=2,4$ and
$6$ (section 2), and next we have calculated the analogous limits for the potentials ${\cal H}_{\rm
CF}(q)=B_{2q}C_{q}^{(2)}+B_{4q}C_{q}^{(4)}+B_{6q}C_{q}^{(6)}$, this time with a fixed $q$ index (section 3).

Collating the results from both the approaches we are able to deduce the ranges $(\Delta E_{min},\Delta
E_{max})$ binding all the crystal fields and the electron states $|J\rangle$ with $1\leq J \leq 8$ (section 4).
The information on the extreme total splittings is valuable and helpful for experimentalists attempting to
assign or verify complicated energy level spectra in crystal-fields.

\section*{2. Dispersion of the total splitting $\Delta E$ of $|J\rangle$ electron states in CFs of individual
             $2^{(k)}$-pole ${\cal H}_{\rm CF}^{(k)}$ for $k=2,4,6$ yielding the same $\sigma^{2}$.}

In order to find the limits of the observed total splittings $\Delta E$ and to compare the splitting capability
of different isomodular ${\cal H}_{\rm CF}^{(k)}$ with respect to $|J\rangle$ states the relevant interaction
matrices of order $2J+1$ for all $1\leq J \leq 8$ have been diagonalized. There were chosen 8, 25 and 30 various
${\cal H}_{\rm CF}^{(k)}$s, both real and virtual, for $k=2,4$ and $6$, respectively.

These intentionally selected representative potentials are characterized by different contributions of the
particular $q$-components. The exact and detailed quantitative compositions of the considered potentials are
comprised in Appendix. The calculated minimal and maximal total splittings, their averages and mean square
deviations are compiled in Tables 1, 2 and 3, respectively for $k=2,4$ and $6$. The complementary CF strength
scale [8], i.e. the spherically averaged absolute values of $|{\cal H}_{\rm CF}^{(k)}|_{\rm av}$, enabled us to
list in a systematic way the different isomodular ${\cal H}_{\rm CF}^{(k)}$s. However, its correlation with
$\Delta E$ is not monotonic and direct.

Our initial assumption comparing the splitting effects of various ${\cal H}_{\rm CF}^{(k)}$s is the constant
second moment $\sigma^{2}$ of the observed splittings instead of the relevant conventional CF strength
[6,7,10,11]. Nevertheless, considering the ${\cal H}_{\rm CF}^{(k)}$ compositions the latter invariant is more
convenient. The energy quantities such as $B_{kq}$, $\Delta E$, $|{\cal H}_{\rm CF}^{(k)}|_{\rm av}$, and
$\delta$ take in these two approaches different values but they are correlated by a simple linear scaling
relationship $\sigma_{k}^{2}=S_{k}^{2}\frac{\left(\langle J||C^{(k)}|| J\rangle\right)^{2}}{(2J+1)(2k+1)}$
[6-9]. Further, for the sake of simplicity, writing the CF Hamiltonians we assume that $S_{k}^{2}=1$ and ignore
the scalar factor $\frac{\sqrt{(2J+1)(2k+1)}}{\langle J||C^{(k)}|| J\rangle}$. The last one ought to be
introduced to ensure the condition $\sigma_{k}^{2}=1$.

However, even the most careful and attentive choice of representative isomodular ${\cal H}_{\rm CF}^{(k)}$ sets
covering the entire range of known calculated $|{\cal H}_{\rm CF}^{(k)}|_{\rm av}$ and preceded by several
surveys, does not guarantee, in general case, to find the exact extreme splittings $\Delta E_{min}$ and $\Delta
E_{max}$. The extreme total splittings proved to be certain have been found only for the quadrupolar ${\cal
H}_{\rm CF}^{(2)}$ acting on all the $|J\rangle$ states, as well as in some particular cases for the higher CF
multipoles (Tables 1-3 vs. Table 10). Since the infinite number of ${\cal H}_{\rm CF}^{(k)}$s cannot be
explicitly inspected, the minimal and maximal total splittings found numerically for the considered ${\cal
H}_{\rm CF}^{(k)}$ sets have been assumed as the extreme ones. Thus, in most cases they have a hypothetical
character of statistical origin due to their sample nature. Fortunately, this intra-multipolar (for a fixed $k$)
$q$-mixing mechanism does not determine the extreme total splittings of the global ${\cal H}_{\rm
CF}=\sum_{k}{\cal H}_{\rm CF}^{(k)}$ (section 3 and 4). The exact numerical approach consisting in mapping of
all the independent CFPs variability ranges would be a generalization of the previous technique employed for the
axial CFs [1]. It is however a hardly realizable task due to its dimensionality even for an individual $k$. Our
more practical choice to overcome this complexity was a numerical experiment that does not guarantee the
absolute character of the found extreme values, although there are some convincing premises making them
reliable. The apparent over-representation of the simple mono-parametric ${\cal H}_{\rm CF}^{(k)}$ in Tables 1-3
leads to the conclusion that the postulated $\Delta E_{min}$ and $\Delta E_{max}$ should be correct or at least
close to the actual extreme values. Still better consistence should hold for the $\Delta E$ averages and the
mean square deviations $\delta$. The CFP mixing within the individual $2^{k}$-poles, i.e. within the component
${\cal H}_{\rm CF}^{(k)}$, generally does not promote the extreme $\Delta E$. The reason for that is the
additivity of $\sigma^{2}$ with respect to $|B_{kq}|^{2}$ (Eq. (2)). It means that every $B_{kq}$ contributes
separately and independently to $\sigma^{2}$ producing however qualitatively various partial splittings.
Considering the linear independence of the involved operators $C_{q}^{(k)}$ for different $q$, which implies
from their orthogonality, the resultant CF effect should come rather from the average of the partial
contributions. However, it does not exclude some particular superpositions corresponding e.g. to
high-symmetrical ${\cal H}_{\rm CF}^{(k)}$s, such as cubic potentials, which promote rather symmetric dichotomic
splittings with $\Delta E$ approaching $2\sigma$. In order to distinguish the statistical data compiled in
Tables 1-3 from the exact results obtained strictly numerically (section 3), the former are given with the
limited accuracy of $10^{-2}\sigma$ whereas the $\overline{\Delta E}$ and $\delta$ with the accuracy of
$10^{-3}\sigma$. A review of the chosen ${\cal H}_{\rm CF}^{(k)}$s with respect to the produced extreme
splittings is discussed below.

In the case of the pure quadrupolar CFs, ${\cal H}_{\rm CF}^{(2)}$, (Table 1 and Appendix) for all the
$|J\rangle$ states with $1\leq J \leq 8$, the $\Delta E_{min}$ is attained under the action of the ${\cal
H}_{\rm CF}^{(2)}=C_{0}^{(2)}$ Hamiltonian (further denoted by the $\left[C_{0}^{(2)}\right]$CF), whereas
$\Delta E_{max}$ either in $\left[C_{2}^{(2)}\right]$ or $\left[C_{1}^{(2)}\right]$CFs. Only for ${\cal H}_{\rm
CF}^{(2)}$s the unambiguous and monotonic relationship between $\Delta E$ and the relevant $\overline{|E_{n}|}$
is observed (Appendix). The $\Delta E_{min}$ corresponds to the largest $\overline{|E_{n}|}$, whereas the
$\Delta E_{max}$ to the smallest $\overline{|E_{n}|}$ (what is understood for the $\sigma^{2}=const$ condition),
and these magnitudes are the nominally extreme ones, indeed. In the quadrupolar case the mean $\overline{\Delta
E}$ for the chosen eight representative ${\cal H}_{\rm CF}^{(2)}$s rises monotonically from $2.343\sigma$ for
$J=1$ up to $3.185\sigma$ for $J=8$ (except the trivial case of $J=3/2$ when it is always $2.0000\sigma$) and is
shifted towards the $\Delta E_{max}$. The mean square deviation $\delta$ is of the order of $0.1\sigma$ and
becomes much smaller within the range $3/2 \leq J \leq 4$ (Table 1). For $J=3/2$ and $2$ the $\Delta E$ is
entirely independent of the ${\cal H}_{\rm CF}^{(2)}$ composition, whereas for $J=5/2$ and $3$ depends only
slightly with the $\delta$ being merely of the order of $0.01\sigma$.

In the case of the $2^{4}$-pole, ${\cal H}_{\rm CF}^{(4)}$, (Table 2 and Appendix) for the noticeable part of
$J$ values $(J=2, 5/2, 11/2\div 8)$ the $\Delta E_{min}$ is achieved in the cubic CF:
$\left[\frac{1}{2}\sqrt{\frac{7}{3}}C_{0}^{(4)}+\frac{1}{2}\sqrt{\frac{5}{3}}C_{4}^{(4)}\right]$ confirming its
tendency to the dichotomic type splittings. In turn, for $J=3$ and $9/2$ the $\Delta E_{min}$ takes place in
$\left[C_{4}^{(4)}\right]$CF, and for $J=7/2$ and $4$ in
$\left[\frac{1}{\sqrt{2}}C_{0}^{(4)}+\frac{1}{\sqrt{2}}C_{2}^{(4)}\right]$CF, although the approximate
magnitudes for both the minima occur in $\left[C_{4}^{(4)}\right]$CF. The case of $J=5$ seems to be quite
intriguing since the $\Delta E$ is completely independent on the ${\cal H}_{\rm CF}^{(4)}$ composition what
means here that $\delta=0$. Further, the $\Delta E_{max}$ for $J=2$ and $5/2$ is gained in
$\left[C_{4}^{(4)}\right]$CF, for $J=3\div 9/2$ in $\left[C_{3}^{(4)}\right]$CF, for $J=11/2\div 13/2$ in
$\left[C_{0}^{(4)}\right]$ axial CF, next for $J=7$ in
$\left[\frac{1}{\sqrt{2}}C_{2}^{(4)}+\frac{1}{\sqrt{2}}C_{3}^{(4)}\right]$CF, and finally for $J=15/2$ and $8$
in $\left[C_{1}^{(4)}\right]$CF. It turns out that the same CF, $\left[C_{0}^{(4)}\right]$, yields $\Delta
E_{min}$ for the states with $J=3\div 9/2$, as well as the $\Delta E_{max}$ for the states with $J=2$ and $5/2$.
It is easy to show on the splitting diagrams and taking into account the relevant crystal-quantum numbers. The
mean $\overline{\Delta E}$ magnitudes do not change considerably with $J$ oscillating around $2.60\sigma$. What
is worthy to notice is the constant $\Delta E=2.3531\sigma$ for $J=5$ and dominating $\overline{\Delta
E}=2.965\sigma$ for $J=3$. In the case of $k=4$ the mean square deviation $\delta$ is generally somewhat higher
than for $k=2$, but also of the order of $0.1\sigma$ and rises apparently for $J=2$ and $3$ decreasing within
the range $9/2 \leq J \leq 6$.

In the case of the $2^{6}$-pole, ${\cal H}_{\rm CF}^{(6)}$, (Table 3 and Appendix) for $J=3$ and $4$ the $\Delta
E_{min}$ has been found in $\left[C_{3}^{(6)}\right]$CF, for $J=7/2$ and $5$ in the cubic CF:
$\left[\frac{1}{2\sqrt{2}}C_{0}^{(6)}-\frac{\sqrt{7}}{2\sqrt{2}}C_{4}^{(6)}\right]$, but in turn for $J=9/2$ in
$\left[\frac{1}{\sqrt{13}}C_{0}^{(6)}+\sqrt{\frac{2}{13}}C_{1}^{(6)}+\sqrt{\frac{2}{13}}C_{2}^{(6)}+
\sqrt{\frac{2}{13}}C_{3}^{(6)}-\sqrt{\frac{2}{13}}C_{4}^{(6)}+\sqrt{\frac{2}{13}}C_{5}^{(6)}+
\sqrt{\frac{2}{13}}C_{6}^{(6)}\right]$CF of rather uniform contribution of all the $q$-components. Its effect is
however close to that in $\left[C_{3}^{(6)}\right]$CF. Finally, for $J=11/2\div 8$ the $\Delta E_{min}$ is
gained in $\left[C_{6}^{(6)}\right]$CF. In turn, the $\Delta E_{max}$ for $J=3$ and $7/2$ is attained in
$\left[C_{6}^{(6)}\right]$CF, and for $J=4$ in $\left[C_{4}^{(6)}\right]$CF. For $J=9/2$ a rather isolated
$\Delta E_{max}$ has been found in
$\left[\frac{1}{\sqrt{3}}C_{1}^{(6)}+\frac{1}{\sqrt{3}}C_{3}^{(6)}+\frac{1}{\sqrt{3}}C_{5}^{(6)}\right]$CF. For
$J=5$ it occurs in $\left[C_{2}^{(6)}\right]$CF, but its effect is almost identical with that in
$\left[C_{3}^{(6)}\right]$ or $\left[C_{5}^{(6)}\right]$CFs. In turn, in the case of $J=6$ it appears in
$\left[\frac{1}{\sqrt{13}}C_{0}^{(6)}+\sqrt{\frac{2}{13}}C_{1}^{(6)}+\sqrt{\frac{2}{13}}C_{2}^{(6)}+
\sqrt{\frac{2}{13}}C_{3}^{(6)}+\sqrt{\frac{2}{13}}C_{4}^{(6)}+\sqrt{\frac{2}{13}}C_{5}^{(6)}-
\sqrt{\frac{2}{13}}C_{6}^{(6)}\right]$CF. At last, for $J=7$ and $15/2$ the $\Delta E_{max}$ occurs in
$\left[\frac{1}{\sqrt{2}}C_{0}^{(6)}+\frac{1}{\sqrt{2}}C_{6}^{(6)}\right]$CF, and finally in the case of $J=8$
it appears in
$\left[\frac{1}{\sqrt{13}}C_{0}^{(6)}-\sqrt{\frac{2}{13}}C_{1}^{(6)}+\sqrt{\frac{2}{13}}C_{2}^{(6)}+
\sqrt{\frac{2}{13}}C_{3}^{(6)}+\sqrt{\frac{2}{13}}C_{4}^{(6)}+\sqrt{\frac{2}{13}}C_{5}^{(6)}+
\sqrt{\frac{2}{13}}C_{6}^{(6)}\right]$CF, which is close to the $\Delta E$ in $\left[C_{2}^{(6)}\right]$CF. The
mean value of $\Delta E$ slightly changes with $J$ oscillating around $2.95\sigma$ and is somewhat enhanced for
$J=4$ and $6$. The mean square deviation fluctuates around $0.2\sigma$ reaching its maximal value
$\delta=0.36\sigma$ for $J=3$, and the minimal value $\delta=0.12\sigma$ for $J=9/2$.

Taking into account the fundamental differences between the three effective $2^{k}$-poles contributing to the
global ${\cal H}_{\rm CF}$ the observed discrepancies in $\Delta E$, $\overline{\Delta E}$, and $\delta$ for the
particular multipoles are surprisingly small with respect to both their magnitudes as well as $J$ dependence
(Tables 1 - 3). Apart from the appreciable increase in $\delta$ roughly in ratio $1:2:3$ for $k=2,4$ and $6$,
respectively, all the three additive contributions can be, in principle, considered as comparable. The greater
dispersion of the $\Delta E$ (it means the greater $\delta$) for higher $k$ corresponds to the higher filling of
the $(2J+1)$-dimensional interaction matrices by the non-zero ${\cal H}_{\rm CF}$ matrix elements.

The observed $\Delta E$ fluctuations result primarily from the arithmetic relations (stemming from the
quantization consequences) between the involved numbers $M_{J}$, $|M_{J}|\leq J$, and $q$, $|q|\leq k$,
according to the definition of the crystal quantum number $\mu$, $M_{J}=\mu ({\rm mod}\; q)$ [2,14]. For
example, the state $|J=3\rangle$ in $\left[C_{6}^{(6)}\right]$CF becomes split into two singlets $(\mu=3$ and
$3)$, whereas all the other substates $|M_{J}\rangle$ remain intact. In consequence, the maximal nominally
admissible total splitting $\Delta E=\Delta {\cal E}=3.7417\sigma$ is attained (Table 10). However, exactly the
same potential splits the state $|J=6\rangle$ into four doublets $(\mu=1,2,4$ and $5$) and five singlets
$(\mu=0,0,0,3$ and $3$) acting on all the initial substates what results in the low value of $\Delta
E=2.6000\sigma$ -- the lowest one found for the state, exceeding however the nominally minimal $\Delta {\cal
E}=2.0059\sigma$. A comparison of the extreme and average total splittings given in Tables 1 - 3 along with the
relevant axial splittings for $k=2,4$ and $6$ presented on the left side of Table 4 is instructive because it
demonstrates the position of the ${\cal H}_{\rm CF}^{(k)}=C_{0}^{(k)}$ among the other ${\cal H}_{\rm CF}^{(k)}$
potentials.

Above we have investigated the effect of the $|J\rangle$ states splitting by the three individual $2^{k}$-poles
separately. Nevertheless, under the action of the global ${\cal H}_{\rm CF}=\sum_{k=2,4,6}{\cal H}_{\rm
CF}^{(k)}$, yielding the same $\sigma^{2}$, there appears the second mechanism effectively leading to the
dispersion of the available $\Delta E$. This is the resultant effect of the $C_{q}^{(k)}$ components with the
same $q$ but different $k$ (three components for $q=0,1$ and $2$, and two ones for $q=3$ and $4$). And this is
the mechanism which turns out to be decisive. Its thorough analysis is consistently provided in the next
section.

\section*{3. Dispersion of the total splitting $\Delta E$ of $|J\rangle$ states  in CFs with fixed $q$:
 ${\cal H}_{\rm CF}(q)=B_{2q}C_{q}^{(2)}+B_{4q}C_{q}^{(4)}+B_{6q}C_{q}^{(6)}$ yielding the same  $\sigma^{2}$.}

For all the ${\cal H}_{\rm CF}^{(k)}$s the splitting capability of the mono-parametric Hamiltonians (with fixed
$|q|$) seems to be dominating or almost dominating. Therefore, in our the next step of the search for the
extreme total splittings we have focused on the $q$-component superpositions ${\cal H}_{\rm
CF}(q)=B_{2q}C_{q}^{(2)}+B_{4q}C_{q}^{(4)}+B_{6q}C_{q}^{(6)}$ with fixed $q$ running from 0 to 4. They have been
optimized with respect to their splitting capability. Obviously, the assumption that the ${\cal H}_{\rm CF}(q)$
is normalized, leading to the same constant $\sigma^{2}$, still remains in power.

In the case of a three-parameter ${\cal H}_{\rm CF}(q)$ form, the method applied previously to the axial CFs [1]
can be adapted here as well. However, because of the non-diagonality of the ${\cal H}_{\rm CF}(q)$ interaction
matrices for $q\neq 0$, the mapping procedure has to be preceded by their diagonalization. The splitting effect
of each $2^{k}$-pole superposition has been analyzed within the three-dimensional spherical reference frame $(R,
\theta, \varphi)$. Taking the radius of the sphere $R=\sigma$, the matrix elements have the form (in $\sigma$):
\begin{eqnarray}
\langle JM_{J}|{\cal H}_{\rm CF}(q)|JM_{J}^{\prime}\rangle
\!&=&\!(-1)^{J\!-\!M_{J}}\!\sqrt{5\!(2J\!+\!1)}\!\left(\begin{array}{ccc}J&2&J\\M_{J}&q&M_{J}^{\prime}
\end{array}\right)\!\sin\theta\!\cos\varphi  \nonumber \\
&+&\;(-1)^{J-M_{J}}3\sqrt{2J+1}\left(\begin{array}{ccc}J&4&J\\M_{J}&q&M_{J}^{\prime}
\end{array}\right)\sin\theta\sin\varphi \nonumber \\
&+&\;(-1)^{J-M_{J}}\sqrt{13(2J+1)}\left(\begin{array}{ccc}J&6&J\\M_{J}&q&M_{J}^{\prime}
\end{array}\right)cos\theta \;\;,
\end{eqnarray}
where the coordinates $0\leq \theta < \pi$ and $0\leq \varphi < 2\pi$ define the multipolar composition of the
${\cal H}_{\rm CF}(q)$. In the case of the two-parameter superpositions for $q=3$ or $4$ the above formula
reduces itself to the simplified form which can be obtained substituting $\varphi=\pi/2$ into Eq.(4) (due to the
lack of the quadrupolar term).

After diagonalization of the matrices, i.e. having the sublevels eigenvalues, and mapping the whole variation
ranges of the coordinates $\theta$ and $\varphi$, the upper and lower limits of the dominating absolute
differences among the eigenvalues at each $(\theta, \varphi)$ point have been calculated numerically. In
consequence, the physically admissible total splittings $\Delta E_{min}$ and $\Delta E_{max}$ of the $|J\rangle$
states in CFs yielding the same $\sigma^{2}$ have been found. The whole ranges of the $\theta$ and $\varphi$
angles digitized with the accuracy of $2\cdot10^{-4}$ and $1\cdot10^{-4}$, respectively, has been swept up
numerically. The evaluation of the upper and lower limits has been facilitated by a short Fortran programme [1].

The calculated $\Delta E_{min}$ and $\Delta E_{max}$ for $q=0\div 4$ are enclosed in Tables 4 - 8, respectively.
To complete these data and make them easy to compare the relevant $\Delta E$ for the solitary contributions to
${\cal H}_{\rm CF}$ (i.e. for $q=5$ or $6$) are appended in Table 9.

\vspace*{0.3cm}

The essential results stand as follows:
\begin{description}
    \item[--] There are three types of the ${\cal H}_{\rm CF}(q)$ with fixed $q$: the first involving all the three
    terms with $k=2,4$ and $6$ for $q=0,1$ and $2$, the second composed of two terms with $k=4$ and $6$ for $q=3$ and
    $4$, and the third based only on the one term with $k=6$ for $q=5$ and $6$. Intuitively, the
    three-multipolar ${\cal H}_{\rm CF}$s are the most effective in achieving the extreme total splittings.
    The results confirm it convincingly. From among all the 30 extreme
    values merely 5 (three $\Delta E_{max}$s and two $\Delta E_{min}$s) come from the two-multipolar ${\cal H}_{\rm
    CF}$s (three for ${\cal H}_{\rm CF}(q=4)$ and two for ${\cal H}_{\rm CF}(q=3)$) (see Tables 4 - 9).
    From among the complete three-multipolar superpositions only ${\cal H}_{\rm CF}(q=1)s$
    lead for  $1 \leq J \leq 8$ (except for $J=4$) to both the largest $\Delta E_{max}$ and the largest $\Delta
    E_{min}$. In turn, the three-multipolar ${\cal H}_{\rm CF}(q=0)s$ lead to the smallest $\Delta E_{min}$
    as well as to the smallest $\Delta E_{max}$ (again except for $J=4$). It confirms unambiguously the
    extreme splitting capabilities of these two particular CF potentials.
    \item[--] There exist ${\cal H}_{\rm CF}(q=1)$s for which the $\Delta E_{max}$ for all the considered
    Kramers states with $J\leq 15/2$ practically reach the maximal, nominally predicted values (compare Table 5 vs.
    Table 10). The observed deviation found for $J=15/2$ amounts to $0.3\%$ ($\Delta E_{max}=3.9880\sigma$
    vs. $\Delta {\cal E}_{max}=4.0000\sigma$), and for $J=13/2$ only $0.1\%$ ($\Delta E_{max}=3.7381\sigma$
    vs. $\Delta {\cal E}_{max}=3.7417\sigma$). These discrepancies follow the characteristic divergence
    between the $\Delta E_{max}$ and $\Delta {\cal E}_{max}$ values appearing for large $J$ values due to
    breaking of the simple model relation $\Delta E_{max}=\sigma\sqrt{2J+1}$. It means that the splitting
    diagrams with two doublets at energies $\pm \Delta E /2$ and all the others at zero energy become unavailable.
    For the ${\cal H}_{\rm CF}(q=2)s$ and the half-integer $J$ the dependence $\Delta E_{max}(J)$ approximates
    closely that for the ${\cal H}_{\rm CF}(q=1)s$ discussed above (Tables 5 and 6).
    \item[--] For the non-Kramers states the $\Delta E_{max}$ is also reached for the majority of $J$
    numbers under the action of the ${\cal H}_{\rm CF}(q=1)s$.
    Here, only three exceptions have been found. For $J=4$ the prevailing
    $\Delta E_{max}=4.0763\sigma$ occurs for the ${\cal H}_{\rm CF}(q=4)$ (Table 8), and in the case of $J=5$
    and $6$ for the ${\cal H}_{\rm CF}(q=3)s$, where $\Delta E_{max}$ attains $3.5904\sigma$ and
    $3.7054\sigma$, respectively. The latter two values only slightly exceed their counterparts for the
    ${\cal H}_{\rm CF}(q=1)s$ (Tables 7 and 5).
    \item[--] The smallest total splittings $\Delta E_{min}$ both for the Kramers and non-Kramers ions have been
    found in the axial CFs for the ${\cal H}_{\rm CF}(q=0)$ (Table 4). Only for $J=15/2$ and $8$ somewhat lower
    $\Delta E_{min}$ have been achieved for the ${\cal H}_{\rm CF}(q=4)$ (Table 8).
\end{description}

\section*{4. Discussion}

The key issue for the estimation of the maximal splittings of electron states in CFs is the fact that the
maximal nominally predicted $\Delta {\cal E}_{max}$ for all the Kramers states with $J\leq 15/2$ are attained
for the properly composed multipolar superpositions in ${\cal H}_{\rm
CF}(q=1)=B_{21}C_{1}^{(2)}+B_{41}C_{1}^{(4)}+B_{61}C_{1}^{(6)}$ (compare Tables 5 and 10). For the Kramers
states the nominally allowed extreme $\Delta {\cal E}$ (resulting solely from the $\sigma^{2}$ constancy
requirement [1,8]) amount to: $\Delta {\cal E}_{max}=\sigma\sqrt{2J+1}$ and $\Delta {\cal E}_{min}=2\sigma$ for
$J=3/2, 7/2, 11/2$ and $15/2$, and $\Delta {\cal E}_{min}=\sigma\frac{2(2J+1)}{\sqrt{(2J+3)(2J-1)}}$ for
$J=5/2,9/2$ and $13/2$. In turn, for the non-Kramers states they are: $\Delta {\cal
E}_{max}=\sigma\sqrt{2(2J+1)}$ and $\Delta {\cal E}_{min}=\sigma\frac{2J+1}{\sqrt{J(J+1)}}$ (see Table 10).

It means, that no other ${\cal H}_{\rm CF}$, of any unrestricted composition, can produce larger splittings.
This fact entitles us to recognize the ${\cal H}_{\rm CF}(q=1)$ as the strongest potential with respect to the
produced total splitting from among all the possible ${\cal H}_{\rm CF}$s yielding the same $\sigma^{2}$. It
confirms also the secondary role of the $q$-mixing mechanism of the $C_{q}^{(k)}$ components in the ${\cal
H}_{\rm CF}^{(k)}$s (Tables 1-3), as well as the $q$ and $k$ cross-mixing mechanism in the ${\cal H}_{\rm CF}$.

Similarly, for the non-Kramers states the ${\cal H}_{\rm CF}(q=1)$ ensures the dominating $\Delta E_{max}$
(Table 5), and the found exceptions for $J=4\div 6$ and $q=3$ and $4$ result presumably from more favourable
configurations of the sublevels for the involved crystal-quantum numbers (section 2). What promotes the large
$\Delta E_{max}$ is putting two of the sublevels (best of all two singlets) possibly far away from the energy
centre of gravity and bringing together the remaining ones in the vicinity of the centre.

The ${\cal H}_{\rm CF}(q=1)$ as the only one does not allow doublets to occur in the energy spectrum (ignoring
an accidental degeneracy) and splits the $|J\rangle$ state into $2J+1$ singlets. In contrary, the ${\cal H}_{\rm
CF}(q=0)$ splits the $|J\rangle$ state always into $J$ doublets and only one singlet $|M_{J}=0\rangle$. The
${\cal H}_{\rm CF}(q)$s for $q=2\div 6$ generate mixtures of doublets and singlets. The splitting diagrams for
the ${\cal H}_{\rm CF}(q\neq 1)$s are symmetrical with respect to their centres of gravity, which means that the
singlet $|M_{J}=0\rangle$ in the non-Kramers states has to always lie at the centre of gravity. This symmetry
does not hold for $q=0$.

In the light of the above results we postulate that the $\Delta E_{max}$ found for the ${\cal H}_{\rm CF}(q=1)s$
completed by the dominating values for the ${\cal H}_{\rm CF}(q=4)$ in the case of $J=4$, as well as for the
${\cal H}_{\rm CF}(q=3)$ in the case of $J=5$ and $6$, are the highest limits of the total CF splittings of the
$|J\rangle$ states for $1\leq J\leq 8$ (Table 11).

The lower limits of the $\Delta E_{min}$ are achievable presumably in the axial CFs of the ${\cal H}_{\rm
CF}(q=0)$. Firstly, these are the smallest values from all the presented in Tables 4 - 9, except slightly
smaller $\Delta E_{min}$s for $J=15/2$ and $8$ under the action of ${\cal H}_{\rm CF}(q=4)$. Secondly, the axial
CFs split the non-Kramers states into the highest number of doublets. This qualitative similarity of the
splitting diagrams of the non-Kramers and Kramers states in the axial CFs must result in the equally small
extreme total splittings (Table 4). Moreover, the obtained $\Delta E_{min}$s do not differ distinctly (contrary
to the $\Delta E_{max}$s for higher $J$) from their nominal limits (Table 10), although the energy distances
should not be treated linearly. The particular weight of the axial CFs results also from their leading role in
the complementary scale of the CF strength [8]. For all the three $2^{k}$-poles the $|{{\cal H}_{\rm
CF}^{(k)}}|_{\rm av}$ values in the axial CFs are the largest in the sets of all the remaining CFs (Appendix).
This suggests rather higher average magnitudes of the absolute energies $\overline{E_{n}}$ of the sublevels and
naturally lower $\Delta E$.

Thus, we postulate that the $\Delta E_{min}$ achieved in the axial CFs of the ${\cal H}_{\rm CF}(q=0)$ for
$1\leq q\leq7$ and in the CFs of the ${\cal H}_{\rm CF}(q=4)$ for $J=15/2$ and $8$ are the lowest possible
limits of the total CF splittings of the $|J\rangle$ states. All the above estimated maximal and minimal total
splittings of the $|J\rangle$ states for $1\leq q\leq 8$ in any CF are compiled in Table 11.

Concluding, two types of the CF Hamiltonians can be distinguished with respect to their extreme splitting
capability. The lower limits of the total splittings are attained in the prevailing number of cases in the axial
CFs of the ${\cal H}_{\rm CF}(q=0)=B_{20}C_{0}^{(2)}+B_{40}C_{0}^{(4)}+B_{60}C_{0}^{(6)}$, whereas the upper
limits in the CFs of the ${\cal H}_{\rm CF}(q=1)=B_{21}C_{1}^{(2)}+B_{41}C_{1}^{(4)}+B_{61}C_{1}^{(6)}$. For the
non-Kramers states in the first case we deal with the maximal number of doublets (equal to $J$), whereas in the
second case with the maximal number of singlets (equal to $2J+1$). In turn, for the Kramers states only doublets
occur. These two particular Hamiltonians composed of the $C_{q}^{(k)}$ operators with the same $q$ (either $0$
or $1$) turn out to be the most effective from the extreme total splittings viewpoint. They contribute to the
same elements of the ${\cal H}_{\rm CF}(q)$ interaction matrices. On the other hand, the components
$C_{q}^{(k)}$ with different $q$ contribute to various matrix elements. This leads to more advanced mixing of
the initial free ion substates which averages the angle distributions of their electron density. Such effect
does not however favour the extreme splittings.

The lower limits of the total splittings $\Delta E_{min}$ lie within the range $(2.0000\sigma,2.4081\sigma )$,
whereas the nominally allowed $\Delta {\cal E}_{min}$ in $(2.0000\sigma,2.1213\sigma )$.

The upper limits $\Delta E_{max}$ can change within the range $(2.0000\sigma,4.1043\sigma )$ compared with the
nominally allowed $\Delta {\cal E}_{max}$ interval $(2.0000\sigma,5.8310\sigma )$. The gap $\Delta
E_{max}-\Delta E_{min}$ varies from $0$ to $1.9961$ (for $J=4$) and is smaller for the Kramers states (Table
11).

It is worthy to remind that all the energy quantities met in the paper are referred to constant $\sigma^{2}$ and
expressed in $\sigma$. However, the $\sigma$ depends not only on the CFPs (Eq.(2)) but in equal degree on the
multipolar characteristics of the electron density distribution of the central ion affected by the CF.

\clearpage




\renewcommand{\baselinestretch}{1.2}
\clearpage
\begin{small}
\begin{table*}[htbp]

{\Large {\bf Appendix}}

\vspace*{0.2cm} \noindent Normalized ($S_{k}^{2}=1$) values of CFPs in the considered ${\cal H}_{\rm CF}^{(k)}$
for $k=2,4$ and $6$ given under the scheme ($B_{k0}, B_{k1}, B_{k-1}, B_{k2}, B_{k-2}, \dots, B_{kk},B_{k-k})$
and the relevant total splitting $\Delta E$ and average absolute value of $E_{n}$, $\overline{|E_{n}|}$, for the
$|J\rangle$ states given under the scheme $\textbf{J}(\Delta E,\overline{|E_{n}|})$. The Hamiltonians are
ordered according to their complementary CF strength $|{{\cal H}_{\rm CF}^{(k)}}|_{\rm av}$ [8].
\begin{center}

\begin{tabular}{llcl}
\hline\hline
No.& ${\cal H}_{\rm CF}^{(2)}$      &              &  $|{{\cal H}_{\rm CF}^{(2)}}|_{\rm av}$  \\
\hline\hline
1 &$(1,0,0,0,0)$        & axial  &  0.385                      \\
&{\footnotesize {\bf 1}(2.1213,0.9427), {\bf 3/2}(2.0000,1.0000), {\bf 2}(2.3905,0.9560), {\bf 5/2}(2.4056,0.8911)} &&\\
&{\footnotesize {\bf 3}(2.5983,0.8247), {\bf 7/2}(2.6184,0.8728), {\bf 4}(2.7356,0.8862), {\bf 9/2}(2.7540,0.8811)}&&\\
&{\footnotesize {\bf 5}(2.8308,0.8647), {\bf 11/2}(2.8443,0.8536), {\bf 6}(2.9008,0.8675), {\bf 13/2}(2.9124,0.8718)}&&\\
&{\footnotesize {\bf 7}(2.9549,0.8686), {\bf 15/2}(2.9641,0.8604), {\bf 8}(2.9973,0.8593)}. &&\\
\hline
2 &$(\frac{1}{2},0,0,\frac{\sqrt{6}}{4},\frac{\sqrt{6}}{4})$        &        &  0.385                      \\
&{\footnotesize {\bf 1}(2.1213,0.9427), {\bf 3/2}(2.0000,1.0000), {\bf 2}(2.3905,0.9560), {\bf 5/2}(2.4056,0.8911)}&&\\
&{\footnotesize {\bf 3}(2.5983,0.8247), {\bf 7/2}(2.6184,0.8728), {\bf 4}(2.7356,0.8862), {\bf 9/2}(2.7540,0.8811)}&&\\
&{\footnotesize {\bf 5}(2.8308,0.8647), {\bf 11/2}(2.8443,0.8536), {\bf 6}(2.9008,0.8675), {\bf 13/2}(2.9124,0.8718)}&&\\
&{\footnotesize {\bf 7}(2.9549,0.8686), {\bf 15/2}(2.9641,0.8604), {\bf 8}(2.9973,0.8593)}. &&\\
\hline
3 &$(\frac{1}{\sqrt{5}},\frac{1}{\sqrt{5}},\frac{1}{\sqrt{5}},-\frac{1}{\sqrt{5}},-\frac{1}{\sqrt{5}})$        &        &  0.381                      \\
&{\footnotesize {\bf 1}(2.3172,0.9253), {\bf 3/2}(2.0000,1.0000), {\bf 2}(2.3905,0.9475), {\bf 5/2}(2.4198,0.8796)}&&\\
&{\footnotesize {\bf 3}(2.6090,0.8283), {\bf 7/2}(2.6633,0.8696), {\bf 4}(2.7759,0.8794), {\bf 9/2}(2.8284,0.8719)}&&\\
&{\footnotesize {\bf 5}(2.9012,0.8543), {\bf 11/2}(2.9481,0.8497), {\bf 6}(2.9992,0.8619), {\bf 13/2}(3.0387,0.8643)}&&\\
&{\footnotesize {\bf 7}(3.0770,0.8591), {\bf 15/2}(3.1081,0.8506), {\bf 8}(3.1383,0.8540)}. &&\\
\hline
4 &$(\frac{1}{\sqrt{5}},\frac{1}{\sqrt{5}},\frac{1}{\sqrt{5}},\frac{1}{\sqrt{5}},\frac{1}{\sqrt{5}})$        &        &  0.374                      \\
&{\footnotesize {\bf 1}(2.4156,0.8834), {\bf 3/2}(2.0000,1.0000), {\bf 2}(2.3905,0.9260), {\bf 5/2}(2.4401,0.8544)}&&\\
&{\footnotesize {\bf 3}(2.6256,0.8830), {\bf 7/2}(2.7158,0.8652), {\bf 4}(2.8255,0.8647), {\bf 9/2}(2.9027,0.8521)}&&\\
&{\footnotesize {\bf 5}(2.9769,0.8351), {\bf 11/2}(3.0364,0.8435), {\bf 6}(3.0903,0.8490), {\bf 13/2}(3.1358,0.8467)}&&\\
&{\footnotesize {\bf 7}(3.1774,0.8392), {\bf 15/2}(3.2137,0.8363), {\bf 8}(3.2462,0.8427)}. &&\\
\hline
5 &$(\frac{1}{\sqrt{3}},\frac{1}{\sqrt{6}},\frac{1}{\sqrt{6}},\frac{1}{\sqrt{6}},\frac{1}{\sqrt{6}})$        &        &  0.373                      \\
&{\footnotesize {\bf 1}(2.4260,0.8737), {\bf 3/2}(2.0000,1.0000), {\bf 2}(2.3905,0.9210), {\bf 5/2}(2.4428,0.8490)}&&\\
&{\footnotesize {\bf 3}(2.6273,0.8336), {\bf 7/2}(2.7221,0.8646), {\bf 4}(2.8322,0.8613), {\bf 9/2}(2.9112,0.8478)}&&\\
&{\footnotesize {\bf 5}(2.9850,0.8343), {\bf 11/2}(3.0465,0.8428), {\bf 6}(3.0999,0.8465), {\bf 13/2}(3.1467,0.8434)}&&\\
&{\footnotesize {\bf 7}(3.1878,0.8348), {\bf 15/2}(3.2244,0.8354), {\bf 8}(3.2573,0.8399)}. &&\\
\hline
6 &$(\frac{1}{\sqrt{5}},\frac{1}{\sqrt{5}}{\rm e}^{i\pi/4},\frac{1}{\sqrt{5}}{\rm e}^{-i\pi/4},\frac{1}{\sqrt{5}},\frac{1}{\sqrt{5}})$        &        &  0.369                      \\
&{\footnotesize {\bf 1}(2.4450,0.8431), {\bf 3/2}(2.0000,1.0000), {\bf 2}(2.3905,0.9055), {\bf 5/2}(2.4483,0.8314)}&&\\
&{\footnotesize {\bf 3}(2.6321,0.8354), {\bf 7/2}(2.7341,0.8627), {\bf 4}(2.8436,0.8519), {\bf 9/2}(2.9267,0.8344)}&&\\
&{\footnotesize {\bf 5}(3.0013,0.8314), {\bf 11/2}(3.0643,0.8404), {\bf 6}(3.1193,0.8385), {\bf 13/2}(3.1668,0.8325)}&&\\
&{\footnotesize {\bf 7}(3.2086,0.8288), {\bf 15/2}(3.2468,0.8327), {\bf 8}(3.2794,0.8334)}. &&\\
\hline
7 &$(0,\frac{1}{\sqrt{2}},\frac{1}{\sqrt{2}},0,0)$        &        &  0.368                      \\
&{\footnotesize {\bf 1}(2.4495,0.8164), {\bf 3/2}(2.0000,1.0000), {\bf 2}(2.3905,0.8920), {\bf 5/2}(2.4495,0.8164)}&&\\
&{\footnotesize {\bf 3}(2.6332,0.8354), {\bf 7/2}(2.7373,0.8627), {\bf 4}(2.8470,0.8439), {\bf 9/2}(2.9310,0.8238)}&&\\
&{\footnotesize {\bf 5}(3.0058,0.8299), {\bf 11/2}(3.0682,0.8404), {\bf 6}(3.1233,0.8328), {\bf 13/2}(3.1709,0.8241)}&&\\
&{\footnotesize {\bf 7}(3.2138,0.8271), {\bf 15/2}(3.2503,0.8327), {\bf 8}(3.2840,0.8288)}.&& \\
\hline
\end{tabular}
\end{center}
\end{table*}
\end{small}
\clearpage

\begin{small}
\begin{table*}[htbp]
{\Large {\bf Appendix} -- cont.}
\begin{center}
\vspace*{0.7cm}

\begin{tabular}{llll}
\hline
No.& ${\cal H}_{\rm CF}^{(2)}$      &              &  $|{{\cal H}_{\rm CF}^{(2)}}|_{\rm av}$  \\
\hline
8 &$(0,0,0,\frac{1}{\sqrt{2}},\frac{1}{\sqrt{2}})$        &        &  0.368                      \\
&{\footnotesize {\bf 1}(2.4495, 0.8164), {\bf 3/2}(2.0000,1.0000), {\bf 2}(2.3905,0.8920), {\bf 5/2}(2.4495,0.8164)}&&\\
&{\footnotesize {\bf 3}(2.6332,0.8354), {\bf 7/2}(2.7373,0.8627), {\bf 4}(2.8470,0.8439), {\bf 9/2}(2.9310,0.8238)}&&\\
&{\footnotesize {\bf 5}(3.0058,0.8299), {\bf 11/2}(3.0682,0.8404), {\bf 6}(3.1233,0.8328), {\bf 13/2}(3.1709,0.8241)}&&\\
&{\footnotesize {\bf 7}(3.2138,0.8271), {\bf 15/2}(3.2503,0.8327), {\bf 8}(3.2840,0.8288)}.&& \\
\hline
\\\\
\hline\hline
No. &${\cal H}_{\rm CF}^{(4)}$      &              &  $|{{\cal H}_{\rm CF}^{(4)}}|_{\rm av}$  \\
\hline\hline
1 &$(1,0,0,0,0,0,0,0,0)$        & axial  &  0.287                      \\
&{\footnotesize {\bf 2}(2.6725,0.8553), {\bf 5/2}(2.3148,1.9259), {\bf 3}(2.7717,0.8525), {\bf 7/2}(2.5074,0.9113)}&&\\
&{\footnotesize {\bf 4}(2.6154,0.9540), {\bf 9/2}(2.3651,0.9221), {\bf 5}(2.3531,0.9273), {\bf 11/2}(2.5544,0.9415)}&&\\
&{\footnotesize {\bf 6}(2.6944,0.9183), {\bf 13/2}(2.7883,0.9103),{\bf 7}(2.8478,0.9144), {\bf 15/2}(2.8812,0.8952)}&&\\
&{\footnotesize {\bf 8}(2.8957,0.8832)}.&& \\
\hline
2 &$(0,\frac{1}{2},\frac{1}{2},\frac{1}{2},\frac{1}{2},0,0,0,0)$         &        &  0.283                      \\
&{\footnotesize {\bf 2}(2.7598,0.8560), {\bf 5/2}(2.3780,0.9061), {\bf 3}(3.0011,0.8326), {\bf 7/2}(2.6228,0.8977)}&&\\
&{\footnotesize {\bf 4}(2.6514,0.9396), {\bf 9/2}(2.4258,0.9164), {\bf 5}(2.3531,0.9273), {\bf 11/2}(2.5222,0.9457)}&&\\
&{\footnotesize {\bf 6}(2.6598,0.9205), {\bf 13/2}(2.7423,0.9081),{\bf 7}(2.8060,0.9086), {\bf 15/2}(2.8416,0.8892)}&&\\
&{\footnotesize {\bf 8}(2.8684,0.8757)}. &&\\
\hline
3 &$(\frac{1}{\sqrt{2}},\frac{1}{2},\frac{1}{2},0,0,0,0,0,0)$        &        &  0.283                      \\
&{\footnotesize {\bf 2}(2.8651,0.8426), {\bf 5/2}(2.3544,0.9149), {\bf 3}(2.7479,0.8429), {\bf 7/2}(2.4981,0.9215)}&&\\
&{\footnotesize {\bf 4}(2.5875,0.9522), {\bf 9/2}(2.3679,0.9193), {\bf 5}(2.3531,0.9233), {\bf 11/2}(2.5472,0.9343)}&&\\
&{\footnotesize {\bf 6}(2.6836,0.9108), {\bf 13/2}(2.7782,0.9014),{\bf 7}(2.8350,0.9005), {\bf 15/2}(2.8848,0.8844)}&&\\
&{\footnotesize {\bf 8}(2.9167,0.8733)}. &&\\
\hline
4 &$(\frac{1}{2}\sqrt{\frac{7}{3}},0,0,0,0,0,0,\frac{1}{2}\sqrt{\frac{5}{6}},\frac{1}{2}\sqrt{\frac{5}{6}})$          &  cubic &  0.280                      \\
&{\footnotesize {\bf 2}(2.0412,0.9801), {\bf 5/2}(2.1213,0.9428), {\bf 3}(2.9312,0.8374), {\bf 7/2}(2.7849,0.7832)}&&\\
&{\footnotesize {\bf 4}(2.7657,0.8874), {\bf 9/2}(2.4467,0.9458), {\bf 5}(2.3531,0.9681), {\bf 11/2}(2.3830,0.9779)}&&\\
&{\footnotesize {\bf 6}(2.5333,0.9746), {\bf 13/2}(2.4549,0.9620),{\bf 7}(2.4516,0.9423), {\bf 15/2}(2.4636,0.9204)}&&\\
&{\footnotesize {\bf 8}(2.5221,0.8980)}. &&\\
\hline
5 &$(\frac{1}{\sqrt{3}},\frac{1}{\sqrt{3}},\frac{1}{\sqrt{3}},0,0,0,0,0,0)$       &        &  0.280                      \\
&{\footnotesize {\bf 2}(2.9268,0.8318), {\bf 5/2}(2.3824,0.9039), {\bf 3}(2.7336,0.8588), {\bf 7/2}(2.5006,0.9291)}&&\\
&{\footnotesize {\bf 4}(2.5650,0.9513), {\bf 9/2}(2.3708,0.9164), {\bf 5}(2.3531,0.9194), {\bf 11/2}(2.5388,0.9280)}&&\\
&{\footnotesize {\bf 6}(2.6728,0.9032), {\bf 13/2}(2.7670,0.8935),{\bf 7}(2.8339,0.8947), {\bf 15/2}(2.8836,0.8760)}&&\\
&{\footnotesize {\bf 8}(2.9340,0.8646)}.&& \\
\hline
6 &$(0,0,0,\frac{1}{2},\frac{1}{2},0,0,\frac{1}{2},\frac{1}{2})$        &        &  0.279                      \\
&{\footnotesize {\bf 2}(2.5981,0.9298), {\bf 5/2}(2.2935,0.9303), {\bf 3}(3.0249,0.8072), {\bf 7/2}(2.6975,0.8477)}&&\\
&{\footnotesize {\bf 4}(2.7594,0.9180), {\bf 9/2}(2.4808,0.9354), {\bf 5}(2.3531,0.9522), {\bf 11/2}(2.4692,0.9623)}&&\\
&{\footnotesize {\bf 6}(2.6155,0.9540), {\bf 13/2}(2.6412,0.9418),{\bf 7}(2.6944,0.9191), {\bf 15/2}(2.7144,0.8976)}&&\\
&{\footnotesize {\bf 8}(2.7472,0.8881)}.&& \\
\hline

\end{tabular}
\end{center}
\end{table*}
\end{small}
\clearpage

\begin{small}
\begin{table*}[htbp]
{\Large {\bf Appendix} -- cont.}
\begin{center}
\vspace*{0.7cm}
\begin{tabular}{llll}
\hline
No. &${\cal H}_{\rm CF}^{(4)}$      &              &  $|{{\cal H}_{\rm CF}^{(4)}}|_{\rm av}$  \\
\hline
7 &$(\frac{1}{\sqrt{2}},0,0,0,0,0,0,\frac{1}{2},\frac{1}{2})$        &        &  0.279                      \\
&{\footnotesize {\bf 2}(2.2358,0.9760), {\bf 5/2}(2.1715,0.9421), {\bf 3}(2.9558,0.8350), {\bf 7/2}(2.7781,0.7815)}&&\\
&{\footnotesize {\bf 4}(2.7918,0.8865), {\bf 9/2}(2.4656,0.9449), {\bf 5}(2.3531,0.9671), {\bf 11/2}(2.3954,0.9758)}&&\\
&{\footnotesize {\bf 6}(2.5419,0.9735), {\bf 13/2}(2.4774,0.9609),{\bf 7}(2.4725,0.9365), {\bf 15/2}(2.4924,0.9192)}&&\\
&{\footnotesize {\bf 8}(2.5382,0.8968)}. &&\\
\hline
8 &$(\frac{1}{3},\frac{1}{3},\frac{1}{3},\frac{1}{3}{\rm e}^{i\pi/2},\frac{1}{3}{\rm e}^{-i\pi/2},\frac{1}{3},\frac{1}{3},\frac{1}{3},\frac{1}{3})$        &        &  0.277                      \\
&{\footnotesize {\bf 2}(2.8047,0.8560), {\bf 5/2}(2.3405,0.9193), {\bf 3}(3.1297,0.8159), {\bf 7/2}(2.7577,0.8095)}&&\\
&{\footnotesize {\bf 4}(2.7846,0.8937), {\bf 9/2}(2.4799,0.9335), {\bf 5}(2.3531,0.9273), {\bf 11/2}(2.4266,0.9675)}&&\\
&{\footnotesize {\bf 6}(2.6220,0.9778), {\bf 13/2}(2.4762,0.9463),{\bf 7}(2.6212,0.9249), {\bf 15/2}(2.6292,0.9012)}&&\\
&{\footnotesize {\bf 8}(2.6780,0.8819)}. &&\\
\hline
9 &$(\frac{1}{\sqrt{2}},0,0,0,0,\frac{1}{2},\frac{1}{2},0,0)$       &        &  0.277                      \\
&{\footnotesize {\bf 2}(2.3385,0.9633), {\bf 5/2}(2.3544,0.9149), {\bf 3}(3.0717,0.8255), {\bf 7/2}(2.7959,0.7637)}&&\\
&{\footnotesize {\bf 4}(2.7450,0.8703), {\bf 9/2}(2.4590,0.9345), {\bf 5}(2.3531,0.9631), {\bf 11/2}(2.3996,0.9738)}&&\\
&{\footnotesize {\bf 6}(2.5441,0.9681), {\bf 13/2}(2.5279,0.9530),{\bf 7}(2.5504,0.9284), {\bf 15/2}(2.5752,0.9084)}&&\\
&{\footnotesize {\bf 8}(2.6248,0.8844)}. &&\\
\hline
10 &$(\frac{1}{3},-\frac{1}{3},-\frac{1}{3},\frac{1}{3},\frac{1}{3},\frac{1}{3},\frac{1}{3},\frac{1}{3},\frac{1}{3})$        &   &  0.276                      \\
&{\footnotesize {\bf 2}(2.7289,0.9103), {\bf 5/2}(2.3736,0.9075), {\bf 3}(3.1662,0.8239), {\bf 7/2}(2.7891,0.7789)}&&\\
&{\footnotesize {\bf 4}(2.7558,0.8766), {\bf 9/2}(2.4751,0.9297), {\bf 5}(2.3531,0.9592), {\bf 11/2}(2.4058,0.9706)}&&\\
&{\footnotesize {\bf 6}(2.6263,0.9908), {\bf 13/2}(2.5357,0.9474),{\bf 7}(2.5794,0.9249), {\bf 15/2}(2.5800,0.9012)}&&\\
&{\footnotesize {\bf 8}(2.6211,0.8782)}. \\
\hline
11 &$(0,0,0,\frac{1}{2},\frac{1}{2},\frac{1}{2},\frac{1}{2},0,0)$        &        &  0.276                     \\
&{\footnotesize {\bf 2}(2.7530,0.8989), {\bf 5/2}(2.4495,0.8164), {\bf 3}(3.0733,0.7874), {\bf 7/2}(2.7314,0.8663)}&&\\
&{\footnotesize {\bf 4}(2.6550,0.8991), {\bf 9/2}(2.4286,0.8975), {\bf 5}(2.3531,0.9423), {\bf 11/2}(2.4630,0.9530)}&&\\
&{\footnotesize {\bf 6}(2.5971,0.9389), {\bf 13/2}(2.6558,0.9126),{\bf 7}(3.0047,0.9028), {\bf 15/2}(2.7684,0.8820)}&&\\
&{\footnotesize {\bf 8}(2.8190,0.8597)}. \\
\hline
12 &$(\frac{1}{3},\frac{1}{3},\frac{1}{3},\frac{1}{3},\frac{1}{3},\frac{1}{3},\frac{1}{3},-\frac{1}{3},-\frac{1}{3})$        &        &  0.274                      \\
&{\footnotesize {\bf 2}(2.5438,0.9284), {\bf 5/2}(2.4110,0.8870), {\bf 3}(3.2860,0.8207), {\bf 7/2}(2.8035,0.7705)}&&\\
&{\footnotesize {\bf 4}(2.7873,0.8667), {\bf 9/2}(2.4789,0.9212), {\bf 5}(2.3531,0.9562), {\bf 11/2}(2.4017,0.9706)}&&\\
&{\footnotesize {\bf 6}(2.6122,0.9876), {\bf 13/2}(2.5312,0.9418),{\bf 7}(2.5980,0.9179), {\bf 15/2}(2.5716,0.8940)}&&\\
&{\footnotesize {\bf 8}(2.6062,0.8708)}. &&\\
\hline
13 &$(\frac{1}{3},\frac{1}{3},\frac{1}{3},\frac{1}{3},\frac{1}{3},-\frac{1}{3}{\rm e}^{i\pi/2},-\frac{1}{3}{\rm e}^{-i\pi/2},\frac{1}{3},\frac{1}{3})$        &        &  0.273                      \\
&{\footnotesize {\bf 2}(2.9228,0.8358), {\bf 5/2}(2.4272,0.8730), {\bf 3}(3.2289,0.8128), {\bf 7/2}(2.7789,0.8171)}&&\\
&{\footnotesize {\bf 4}(2.7612,0.8838), {\bf 9/2}(2.4761,0.9136), {\bf 5}(2.3531,0.9482), {\bf 11/2}(2.4224,0.9644)}&&\\
&{\footnotesize {\bf 6}(2.6317,0.9767), {\bf 13/2}(2.5705,0.9306),{\bf 7}(2.6352,0.9063), {\bf 15/2}(2.6436,0.8832)}&&\\
&{\footnotesize {\bf 8}(2.6928,0.8671)}. &&\\
\hline
\end{tabular}
\end{center}
\end{table*}
\end{small}
\clearpage

\begin{small}
\begin{table*}[htbp]
{\Large {\bf Appendix} -- cont.}
\begin{center}
\vspace*{0.7cm}
\begin{tabular}{llll}
\hline
No. &${\cal H}_{\rm CF}^{(4)}$      &              &  $|{{\cal H}_{\rm CF}^{(4)}}|_{\rm av}$  \\
\hline
14 &$(0,\frac{1}{2},\frac{1}{2},0,0,0,0,\frac{1}{2},\frac{1}{2})$          &        &  0.273                      \\
&{\footnotesize {\bf 2}(2.4961,0.8996), {\bf 5/2}(2.4495,0.8164), {\bf 3}(3.3225,0.7921), {\bf 7/2}(2.8044,0.7934)}&&\\
&{\footnotesize {\bf 4}(2.7774,0.8676), {\bf 9/2}(2.4837,0.8918), {\bf 5}(2.3531,0.9423), {\bf 11/2}(2.4287,0.9675)}&&\\
&{\footnotesize {\bf 6}(2.5744,0.9465), {\bf 13/2}(2.5862,0.9182),{\bf 7}(2.6538,0.9016), {\bf 15/2}(2.6640,0.8880)}&&\\
&{\footnotesize {\bf 8}(2.7064,0.8659)}. &&\\
\hline
15 &$(\frac{3}{2\sqrt{11}},0,0,0,0,0,0,\frac{1}{2}\sqrt{\frac{35}{22}},\frac{1}{2}\sqrt{\frac{35}{22}})$         & plane square       &  0.271                      \\
&{\footnotesize {\bf 2}(2.8208,0.9023), {\bf 5/2}(2.3398,0.9193), {\bf 3}(2.8701,0.9048), {\bf 7/2}(2.6109,0.8935)}&&\\
&{\footnotesize {\bf 4}(2.7198,0.9324), {\bf 9/2}(2.4542,0.9307), {\bf 5}(2.3531,0.9432), {\bf 11/2}(2.4463,0.9447)}&&\\
&{\footnotesize {\bf 6}(2.5798,0.9346), {\bf 13/2}(2.5829,0.9216),{\bf 7}(2.6480,0.8993), {\bf 15/2}(2.6676,0.8772)}&&\\
&{\footnotesize {\bf 8}(2.7039,0.8572)}.&& \\
\hline
16 &$(0,0,0,\frac{1}{\sqrt{2}},\frac{1}{\sqrt{2}},0,0,0,0)$        &        &  0.269                      \\
&{\footnotesize {\bf 2}(2.3908,0.8922), {\bf 5/2}(2.4495,0.8164), {\bf 3}(3.3654,0.7945), {\bf 7/2}(2.8222,0.7526)}&&\\
&{\footnotesize {\bf 4}(2.7999,0.8424), {\bf 9/2}(2.4922,0.8880), {\bf 5}(2.3531,0.9393), {\bf 11/2}(2.4141,0.9706)}&&\\
&{\footnotesize {\bf 6}(2.5657,0.9454), {\bf 13/2}(2.5537,0.9171),{\bf 7}(2.6224,0.9016), {\bf 15/2}(2.6016,0.8856)}&&\\
&{\footnotesize {\bf 8}(2.6260,0.8634)}.&& \\
\hline
17 &$(0,\frac{1}{2},\frac{1}{2},0,0,\frac{1}{2},\frac{1}{2},0,0)$          &        &  0.268                      \\
&{\footnotesize {\bf 2}(2.2922,0.8942), {\bf 5/2}(2.4495,0.8164), {\bf 3}(3.3821,0.8048), {\bf 7/2}(2.8281,0.7238)}&&\\
&{\footnotesize {\bf 4}(2.8152,0.8226), {\bf 9/2}(2.4941,0.8880), {\bf 5}(2.3531,0.9403), {\bf 11/2}(2.4058,0.9717)}&&\\
&{\footnotesize {\bf 6}(2.5365,0.9465), {\bf 13/2}(2.5312,0.9171),{\bf 7}(2.9710,0.8888), {\bf 15/2}(2.5680,0.8700)}&&\\
&{\footnotesize {\bf 8}(2.5901,0.8535)}. &&\\
\hline
18 &$(0,0,0,0,0,\frac{1}{\sqrt{2}},\frac{1}{\sqrt{2}},0,0)$      &   &  0.266                     \\
&{\footnotesize {\bf 2}(2.2358,0.8942), {\bf 5/2}(2.4495,0.8164), {\bf 3}(3.3844,0.8056), {\bf 7/2}(2.8281,0.7068)}&&\\
&{\footnotesize {\bf 4}(2.8161,0.8118), {\bf 9/2}(2.4950,0.8880), {\bf 5}(2.3531,0.9403), {\bf 11/2}(2.4037,0.9717)}&&\\
&{\footnotesize {\bf 6}(2.5343,0.9465), {\bf 13/2}(2.5279,0.9171),{\bf 7}(2.6131,0.8888), {\bf 15/2}(2.5620,0.8616)}&&\\
&{\footnotesize {\bf 8}(2.5827,0.8535)}. &&\\
\hline
19 &$(0,\frac{1}{\sqrt{2}},\frac{1}{\sqrt{2}},0,0,0,0,0,0)$        &        &  0.266                      \\
&{\footnotesize {\bf 2}(3.0475,0.7788), {\bf 5/2}(2.4495,0.8164), {\bf 3}(2.7114,0.8699), {\bf 7/2}(2.5312,0.9487)}&&\\
&{\footnotesize {\bf 4}(2.5101,0.9270), {\bf 9/2}(2.3802,0.8927), {\bf 5}(2.3531,0.9015), {\bf 11/2}(2.5014,0.9062)}&&\\
&{\footnotesize {\bf 6}(2.6230,0.8794), {\bf 13/2}(2.7176,0.8565),{\bf 7}(2.8071,0.8447), {\bf 15/2}(2.8956,0.8268)}&&\\
&{\footnotesize {\bf 8}(2.9822,0.8127)}.&& \\
\hline
20 &$(\frac{1}{3},\frac{1}{3},\frac{1}{3},\frac{1}{3},\frac{1}{3},\frac{1}{3},\frac{1}{3},\frac{1}{3},\frac{1}{3})$        &        &  0.265                      \\
&{\footnotesize {\bf 2}(2.6035,0.9090), {\bf 5/2}(2.4478,0.8333), {\bf 3}(2.7645,0.8977), {\bf 7/2}(2.4802,0.9588)}&\\
&{\footnotesize {\bf 4}(2.4714,0.9351), {\bf 9/2}(2.3613,0.8984), {\bf 5}(2.3531,0.8975), {\bf 11/2}(2.4692,0.8917)}&\\
&{\footnotesize {\bf 6}(2.6890,0.8956), {\bf 13/2}(2.6513,0.8475),{\bf 7}(2.7305,0.8389), {\bf 15/2}(2.7840,0.8316)}&\\
&{\footnotesize {\bf 8}(2.8474,0.8226)}. \\
\hline
\end{tabular}
\end{center}
\end{table*}
\end{small}
\clearpage

\begin{small}
\begin{table*}[htbp]
{\Large {\bf Appendix} -- cont.}
\begin{center}
\vspace*{0.7cm}
\begin{tabular}{llll}
\hline
No. &${\cal H}_{\rm CF}^{(4)}$      &              &  $|{{\cal H}_{\rm CF}^{(4)}}|_{\rm av}$  \\
\hline
21 &$(0,0,0,0,0,\frac{1}{2},\frac{1}{2},\frac{1}{2},\frac{1}{2})$         &        &  0.264                     \\
&{\footnotesize {\bf 2}(3.0542,0.7748), {\bf 5/2}(2.4495,0.8164), {\bf 3}(2.8169,0.8977), {\bf 7/2}(2.5227,0.9504)}&&\\
&{\footnotesize {\bf 4}(2.5326,0.9261), {\bf 9/2}(2.3765,0.8927), {\bf 5}(2.3531,0.9005), {\bf 11/2}(2.4827,0.9021)}&&\\
&{\footnotesize {\bf 6}(2.6014,0.8783), {\bf 13/2}(2.6794,0.8576),{\bf 7}(2.7595,0.8377), {\bf 15/2}(2.8284,0.8196)}&&\\
&{\footnotesize {\bf 8}(2.8994,0.8151)}. &&\\
\hline
22 &$(\frac{1}{\sqrt{3}},0,0,\frac{1}{\sqrt{3}},\frac{1}{\sqrt{3}},0,0,0,0)$          &        &  0.263                     \\
&{\footnotesize {\bf 2}(3.0623,0.7929), {\bf 5/2}(2.4228,0.8774), {\bf 3}(2.7606,0.9334), {\bf 7/2}(2.4149,0.9554)}&&\\
&{\footnotesize {\bf 4}(2.5056,0.9477), {\bf 9/2}(2.3736,0.9126), {\bf 5}(2.3531,0.9064), {\bf 11/2}(2.4723,0.8969)}&&\\
&{\footnotesize {\bf 6}(2.5895,0.8772), {\bf 13/2}(2.6424,0.8632),{\bf 7}(2.7200,0.8505), {\bf 15/2}(2.7708,0.8376)}&&\\
&{\footnotesize {\bf 8}(2.8350,0.8287)}. &&\\
\hline
23 &$(\frac{1}{3},\frac{1}{3}{\rm e}^{i\pi/4},\frac{1}{3}{\rm e}^{-i\pi/4},\frac{1}{3},\frac{1}{3},\frac{1}{3},\frac{1}{3},\frac{1}{3},\frac{1}{3})$       &        &  0.257                      \\
&{\footnotesize {\bf 2}(2.7779,0.8573), {\bf 5/2}(2.4495,0.8164), {\bf 3}(2.5748,0.8969), {\bf 7/2}(2.4353,0.9682)}&&\\
&{\footnotesize {\bf 4}(2.4300,0.9342), {\bf 9/2}(2.3395,0.8937), {\bf 5}(2.3531,0.8895), {\bf 11/2}(2.4910,0.8771)}&&\\
&{\footnotesize {\bf 6}(2.6717,0.8805), {\bf 13/2}(2.6929,0.8295),{\bf 7}(2.7804,0.8215), {\bf 15/2}(2.7660,0.8136)}&&\\
&{\footnotesize {\bf 8}(2.9464,0.7966)}. &&\\
\hline
24 &$(\frac{1}{\sqrt{2}},0,0,\frac{1}{2},\frac{1}{2},0,0,0,0)$         &        &  0.257                     \\
&{\footnotesize {\bf 2}(3.1334,0.7245), {\bf 5/2}(2.4412,0.8532), {\bf 3}(2.5336,0.9342), {\bf 7/2}(2.3326,0.9750)}&&\\
&{\footnotesize {\bf 4}(2.3625,0.9486), {\bf 9/2}(2.3138,0.9050), {\bf 5}(2.3531,0.8835), {\bf 11/2}(2.4806,0.8626)}&&\\
&{\footnotesize {\bf 6}(2.5938,0.8372), {\bf 13/2}(2.6570,0.8284),{\bf 7}(2.7363,0.8203), {\bf 15/2}(2.7972,0.8160)}&&\\
&{\footnotesize {\bf 8}(2.8660,0.8139)}. &&\\
\hline
25 &$(0,0,0,0,0,0,0,\frac{1}{\sqrt{2}},\frac{1}{\sqrt{2}})$         &        &  0.251                      \\
&{\footnotesize {\bf 2}(3.1623,0.6326), {\bf 5/2}(2.4495,0.8164), {\bf 3}(2.5225,0.9183), {\bf 7/2}(2.3351,0.9826)}&&\\
&{\footnotesize {\bf 4}(2.3796,0.9396), {\bf 9/2}(2.2901,0.8946), {\bf 5}(2.3531,0.8527), {\bf 11/2}(2.4838,0.8168)}&&\\
&{\footnotesize {\bf 6}(2.5960,0.8004), {\bf 13/2}(2.6614,0.7195),{\bf 7}(2.7386,0.7971), {\bf 15/2}(2.8020,0.8028)}&&\\
&{\footnotesize {\bf 8}(2.8684,0.7991)}. &&\\
\hline
\\\\\\
\hline\hline
No. &${\cal H}_{\rm CF}^{(6)}$      &              &  $|{{\cal H}_{\rm CF}^{(6)}}|_{\rm av}$  \\
\hline\hline
1 &$(1,0,0,0,0,0,0,0,0,0,0,0,0)$        & axial  &  0.239                      \\
&{\footnotesize {\bf 3}(3.0459,0.7956), {\bf 7/2}(2.4373,0.8709), {\bf 4}(2.8318,0.8091), {\bf 9/2}(2.5848,0.9361)}&&\\
&{\footnotesize {\bf 5}(2.6296,0.9112), {\bf 11/2}(2.8952,0.8793), {\bf 6}(2.9640,0.8840), {\bf 13/2}(2.8965,0.8418)}&&\\
&{\footnotesize {\bf 7}(2.8655,0.8951), {\bf 15/2}(2.9306,0.9201),{\bf 8}(2.9004,0.9009)}.&&\\
\hline
2 &$(\frac{1}{\sqrt{7}},0,0,\frac{1}{\sqrt{7}},\frac{1}{\sqrt{7}},0,0,\frac{1}{\sqrt{7}},\frac{1}{\sqrt{7}},0,0,\frac{1}{\sqrt{7}},\frac{1}{\sqrt{7}})$        &        &  0.236                      \\
&{\footnotesize {\bf 3}(3.2615,0.7832), {\bf 7/2}(2.6117,0.8597), {\bf 4}(3.0330,0.8231), {\bf 9/2}(2.5905,0.9315)}&&\\
&{\footnotesize {\bf 5}(2.6260,0.9052), {\bf 11/2}(3.0763,0.8243), {\bf 6}(2.9250,0.8723), {\bf 13/2}(2.8627,0.8526)}&&\\
&{\footnotesize {\bf 7}(2.8431,0.8979), {\bf 15/2}(2.9032,0.9173),{\bf 8}(2.8751,0.8994)}.&&\\
\hline

\end{tabular}
\end{center}
\end{table*}
\end{small}
\clearpage

\begin{small}
\begin{table*}[htbp]
{\Large {\bf Appendix} -- cont.}
\begin{center}
\vspace*{0.7cm}
\begin{tabular}{llll}
\hline
No. &${\cal H}_{\rm CF}^{(6)}$      &              &  $|{{\cal H}_{\rm CF}^{(6)}}|_{\rm av}$  \\
\hline
3 &$(\frac{1}{\sqrt{2}},\frac{1}{2},\frac{1}{2},0,0,0,0,0,0,0,0,0,0)$        &        &  0.235                      \\
&{\footnotesize {\bf 3}(3.2672,0.7832), {\bf 7/2}(2.6229,0.8587), {\bf 4}(2.8286,0.8275), {\bf 9/2}(2.5893,0.9304)}&&\\
&{\footnotesize {\bf 5}(2.6201,0.9040), {\bf 11/2}(2.8602,0.8706), {\bf 6}(2.9289,0.8723), {\bf 13/2}(2.8708,0.8526)}&&\\
&{\footnotesize {\bf 7}(2.8375,0.8979), {\bf 15/2}(2.8974,0.9173), {\bf 8}(2.8706,0.8979)}.&&\\
\hline
4 &$(0,\frac{1}{2},\frac{1}{2},\frac{1}{2},\frac{1}{2},0,0,0,0,0,0,0,0)$        &        &  0.234                   \\
&{\footnotesize {\bf 3}(3.1995,0.7870), {\bf 7/2}(2.6352,0.8689), {\bf 4}(3.0806,0.8448), {\bf 9/2}(2.6520,0.9372)}&&\\
&{\footnotesize {\bf 5}(2.8329,0.9100), {\bf 11/2}(2.9164,0.8543), {\bf 6}(2.9906,0.8645), {\bf 13/2}(2.9410,0.8418)}&&\\
&{\footnotesize {\bf 7}(2.9213,0.8881), {\bf 15/2}(2.9465,0.9115), {\bf 8}(2.9286,0.9053)}.&&\\
\hline
5 &$(0,0,0,\frac{1}{\sqrt{6}},\frac{1}{\sqrt{6}},0,0,\frac{1}{\sqrt{6}},\frac{1}{\sqrt{6}},0,0,\frac{1}{\sqrt{6}},\frac{1}{\sqrt{6}})$        &        &  0.233                      \\
&{\footnotesize {\bf 3}(2.9047,0.8738), {\bf 7/2}(2.4934,0.9137), {\bf 4}(3.1595,0.7810), {\bf 9/2}(2.6019,0.9224)}&&\\
&{\footnotesize {\bf 5}(2.8544,0.9136), {\bf 11/2}(2.9889,0.8431), {\bf 6}(3.3176,0.8437), {\bf 13/2}(3.1150,0.8607)}&&\\
&{\footnotesize {\bf 7}(3.2495,0.8532), {\bf 15/2}(3.0618,0.8999), {\bf 8}(2.9613,0.9053)}.&&\\
\hline
6 &$(\frac{1}{\sqrt{13}},\frac{1}{\sqrt{13}},\frac{1}{\sqrt{13}},\frac{1}{\sqrt{13}},\frac{1}{\sqrt{13}},\frac{1}{\sqrt{13}},\frac{1}{\sqrt{13}},-\frac{1}{\sqrt{13}},-\frac{1}{\sqrt{13}},\frac{1}{\sqrt{13}},\frac{1}{\sqrt{13}},\frac{1}{\sqrt{13}},\frac{1}{\sqrt{13}})$        &        &  0.231                      \\
&{\footnotesize {\bf 3}(2.4459,0.9740), {\bf 7/2}(2.5679,0.8362), {\bf 4}(2.6869,0.9616), {\bf 9/2}(2.3134,0.9657)}&&\\
&{\footnotesize {\bf 5}(2.7384,0.8287), {\bf 11/2}(2.9864,0.8481), {\bf 6}(3.4606,0.8762), {\bf 13/2}(3.1663,0.8715)}&&\\
&{\footnotesize {\bf 7}(3.0372,0.8323), {\bf 15/2}(2.8397,0.8019), {\bf 8}(2.9360,0.8310)}.&&\\
\hline
7 &$(0,0,0,\frac{1}{2},\frac{1}{2},\frac{1}{2},\frac{1}{2},0,0,0,0,0,0)$         &        &  0.231                     \\
&{\footnotesize {\bf 3}(3.0097,0.8376), {\bf 7/2}(2.7555,0.8485), {\bf 4}(3.1011,0.8826), {\bf 9/2}(2.5825,0.9087)}&&\\
&{\footnotesize {\bf 5}(2.8628,0.8598), {\bf 11/2}(3.0251,0.8268), {\bf 6}(3.2591,0.8411), {\bf 13/2}(3.1231,0.8216)}&&\\
&{\footnotesize {\bf 7}(3.1350,0.8393), {\bf 15/2}(2.9840,0.8826), {\bf 8}(3.0148,0.8726)}.&&\\
\hline
8 &$(0,0,0,\frac{1}{2},\frac{1}{2},0,0,0,0,0,0,\frac{1}{2},\frac{1}{2})$        &        &  0.230                      \\
&{\footnotesize {\bf 3}(2.8456,0.8957), {\bf 7/2}(2.5587,0.9413), {\bf 4}(3.3175,0.7258), {\bf 9/2}(2.6714,0.8916)}&&\\
&{\footnotesize {\bf 5}(2.8975,0.8849), {\bf 11/2}(3.0576,0.7856), {\bf 6}(3.4697,0.8307), {\bf 13/2}(3.2391,0.8364)}&&\\
&{\footnotesize {\bf 7}(3.3109,0.8295), {\bf 15/2}(3.1238,0.8610), {\bf 8}(2.9509,0.8771)}.&&\\
\hline
9 &$(\frac{1}{\sqrt{2}},0,0,0,0,0,0,0,0,0,0,\frac{1}{2},\frac{1}{2})$         &   &  0.229                      \\
&{\footnotesize {\bf 3}(2.6462,0.9234), {\bf 7/2}(2.4200,0.9637), {\bf 4}(3.3402,0.7377), {\bf 9/2}(2.6156,0.9133)}&&\\
&{\footnotesize {\bf 5}(2.9692,0.9148), {\bf 11/2}(3.0563,0.8668), {\bf 6}(3.3670,0.8112), {\bf 13/2}(3.2432,0.8486)}&&\\
&{\footnotesize {\bf 7}(3.4366,0.8462), {\bf 15/2}(3.1354,0.8624), {\bf 8}(3.0193,0.8949)}.&&\\
\hline
10 &$(\frac{1}{\sqrt{13}},\frac{1}{\sqrt{13}},\frac{1}{\sqrt{13}},-\frac{1}{\sqrt{13}},-\frac{1}{\sqrt{13}},\frac{1}{\sqrt{13}},\frac{1}{\sqrt{13}},\frac{1}{\sqrt{13}},\frac{1}{\sqrt{13}},\frac{1}{\sqrt{13}},\frac{1}{\sqrt{13}},\frac{1}{\sqrt{13}},\frac{1}{\sqrt{13}})$        &        &  0.228                      \\
&{\footnotesize {\bf 3}(2.7273,0.8948), {\bf 7/2}(2.5597,0.9403), {\bf 4}(3.4191,0.8015), {\bf 9/2}(2.6042,0.8973)}&&\\
&{\footnotesize {\bf 5}(3.0625,0.9088), {\bf 11/2}(3.0713,0.8755), {\bf 6}(3.6491,0.8879), {\bf 13/2}(3.2067,0.8337)}&&\\
&{\footnotesize {\bf 7}(3.3444,0.8379), {\bf 15/2}(3.1065,0.8639), {\bf 8}(3.0446,0.8920)}.&&\\
\hline
11 &$(\frac{1}{\sqrt{13}},\frac{1}{\sqrt{13}},\frac{1}{\sqrt{13}},-\frac{1}{\sqrt{13}},-\frac{1}{\sqrt{13}},\frac{1}{\sqrt{13}},\frac{1}{\sqrt{13}},-\frac{1}{\sqrt{13}},-\frac{1}{\sqrt{13}},\frac{1}{\sqrt{13}},\frac{1}{\sqrt{13}},\frac{1}{\sqrt{13}},\frac{1}{\sqrt{13}})$        &        &  0.227                      \\
&{\footnotesize {\bf 3}(2.8303,0.9206), {\bf 7/2}(2.7494,0.8536), {\bf 4}(2.9724,0.8426), {\bf 9/2}(2.4936,0.8939)}&&\\
&{\footnotesize {\bf 5}(2.8999,0.8490), {\bf 11/2}(3.1100,0.8143), {\bf 6}(3.6387,0.8866), {\bf 13/2}(3.2432,0.8364)}&&\\
&{\footnotesize {\bf 7}(3.2132,0.8127), {\bf 15/2}(2.9984,0.8480), {\bf 8}(2.8840,0.8697)}.&&\\
\hline

\end{tabular}
\end{center}
\end{table*}
\end{small}
\clearpage

\begin{small}
\begin{table*}[htbp]
{\Large {\bf Appendix} -- cont.}
\begin{center}
\vspace*{0.7cm}
\begin{tabular}{llll}
\hline
No. &${\cal H}_{\rm CF}^{(6)}$      &              &  $|{{\cal H}_{\rm CF}^{(6)}}|_{\rm av}$  \\
\hline
12 &$(\frac{1}{\sqrt{3}},0,0,0,0,0,0,0,0,0,0,\frac{1}{\sqrt{3}},\frac{1}{\sqrt{3}})$         &        &  0.226                     \\
&{\footnotesize {\bf 3}(3.0555,0.8814), {\bf 7/2}(2.3874,0.9484), {\bf 4}(3.3553,0.7853), {\bf 9/2}(2.6258,0.9247)}&&\\
&{\footnotesize {\bf 5}(2.9489,0.9339), {\bf 11/2}(2.9889,0.9005), {\bf 6}(3.2487,0.8515), {\bf 13/2}(3.1285,0.8688)}&&\\
&{\footnotesize {\bf 7}(3.3305,0.8686), {\bf 15/2}(3.0763,0.8999), {\bf 8}(3.0312,0.9143)}.&&\\
\hline
13 &$(-\frac{1}{\sqrt{13}},\frac{1}{\sqrt{13}},\frac{1}{\sqrt{13}},\frac{1}{\sqrt{13}},\frac{1}{\sqrt{13}},\frac{1}{\sqrt{13}},\frac{1}{\sqrt{13}},\frac{1}{\sqrt{13}},\frac{1}{\sqrt{13}},\frac{1}{\sqrt{13}},\frac{1}{\sqrt{13}},\frac{1}{\sqrt{13}},\frac{1}{\sqrt{13}})$        &        &  0.225                      \\
&{\footnotesize {\bf 3}(3.3226,0.7937), {\bf 7/2}(2.7504,0.8505), {\bf 4}(3.1325,0.8653), {\bf 9/2}(2.5643,0.8871)}&&\\
&{\footnotesize {\bf 5}(2.8568,0.8909), {\bf 11/2}(2.9264,0.8743), {\bf 6}(3.5243,0.9048), {\bf 13/2}(3.0071,0.8567)}&&\\
&{\footnotesize {\bf 7}(2.9870,0.8770), {\bf 15/2}(2.9090,0.8855), {\bf 8}(2.8974,0.8786)}.&&\\
\hline
14 &$(\frac{1}{\sqrt{5}},0,0,0,0,\frac{1}{\sqrt{5}},\frac{1}{\sqrt{5}},0,0,0,0,\frac{1}{\sqrt{5}},\frac{1}{\sqrt{5}})$        &        &  0.224                      \\
&{\footnotesize {\bf 3}(2.7197,0.9406), {\bf 7/2}(2.6790,0.8923), {\bf 4}(3.3737,0.7615), {\bf 9/2}(2.5563,0.9087)}&&\\
&{\footnotesize {\bf 5}(2.6906,0.8861), {\bf 11/2}(3.0001,0.8331), {\bf 6}(3.2981,0.8710), {\bf 13/2}(3.2459,0.8594)}&&\\
&{\footnotesize {\bf 7}(3.2537,0.8518), {\bf 15/2}(2.9998,0.8596), {\bf 8}(2.8186,0.8786)}.&&\\
\hline
15 &$(\frac{1}{\sqrt{13}},-\frac{1}{\sqrt{13}},-\frac{1}{\sqrt{13}},\frac{1}{\sqrt{13}},\frac{1}{\sqrt{13}},\frac{1}{\sqrt{13}},\frac{1}{\sqrt{13}},\frac{1}{\sqrt{13}},\frac{1}{\sqrt{13}},\frac{1}{\sqrt{13}},\frac{1}{\sqrt{13}},\frac{1}{\sqrt{13}},\frac{1}{\sqrt{13}})$        &        &  0.223                      \\
&{\footnotesize {\bf 3}(3.1442,0.8347), {\bf 7/2}(2.6444,0.9056), {\bf 4}(3.3315,0.8307), {\bf 9/2}(2.6042,0.9076)}&&\\
&{\footnotesize {\bf 5}(3.0314,0.9124), {\bf 11/2}(2.9789,0.8843), {\bf 6}(3.5178,0.8892), {\bf 13/2}(3.0907,0.8513)}&&\\
&{\footnotesize {\bf 7}(3.2048,0.8630), {\bf 15/2}(3.0301,0.8841), {\bf 8}(3.0802,0.8860)}.&&\\
\hline
16 &$(\frac{1}{2\sqrt{2}},0,0,0,0,0,0,-\frac{\sqrt{7}}{4},-\frac{\sqrt{7}}{4},0,0,0,0)$          &  cubic      &  0.223                      \\
&{\footnotesize {\bf 3}(2.5852,0.9492), {\bf 7/2}(2.2160,0.9851), {\bf 4}(3.4321,0.7420), {\bf 9/2}(2.5437,0.9201)}&&\\
&{\footnotesize {\bf 5}(2.6033,0.8885), {\bf 11/2}(2.8565,0.8318), {\bf 6}(3.2929,0.8099), {\bf 13/2}(3.2486,0.8567)}&&\\
&{\footnotesize {\bf 7}(3.2565,0.8588), {\bf 15/2}(2.9782,0.8596), {\bf 8}(2.7889,0.8816)}.&&\\
\hline
17 &$(\frac{1}{\sqrt{2}},0,0,0,0,0,0,\frac{1}{2},\frac{1}{2},0,0,0,0)$       &  &  0.223                      \\
&{\footnotesize {\bf 3}(2.6023,0.8910), {\bf 7/2}(2.4271,0.9668), {\bf 4}(3.5879,0.7669), {\bf 9/2}(2.6498,0.8882)}&&\\
&{\footnotesize {\bf 5}(3.0745,0.9017), {\bf 11/2}(3.0488,0.8755), {\bf 6}(3.3228,0.8112), {\bf 13/2}(3.1960,0.8297)}&&\\
&{\footnotesize {\bf 7}(3.2565,0.8113), {\bf 15/2}(3.1065,0.8437), {\bf 8}(3.0208,0.8771)}.&&\\
\hline
18 &$(\frac{1}{\sqrt{13}},\frac{1}{\sqrt{13}},\frac{1}{\sqrt{13}},\frac{1}{\sqrt{13}},\frac{1}{\sqrt{13}},\frac{1}{\sqrt{13}},\frac{1}{\sqrt{13}},\frac{1}{\sqrt{13}},\frac{1}{\sqrt{13}},\frac{1}{\sqrt{13}},\frac{1}{\sqrt{13}},-\frac{1}{\sqrt{13}},-\frac{1}{\sqrt{13}})$        &        &  0.222                      \\
&{\footnotesize {\bf 3}(3.1490,0.8347), {\bf 7/2}(2.7055,0.8811), {\bf 4}(3.2731,0.8545), {\bf 9/2}(2.5814,0.8825)}&&\\
&{\footnotesize {\bf 5}(3.0434,0.8993), {\bf 11/2}(3.0113,0.8868), {\bf 6}(3.6894,0.8658), {\bf 13/2}(3.0934,0.8472)}&&\\
&{\footnotesize {\bf 7}(3.1503,0.8588), {\bf 15/2}(2.9984,0.8798), {\bf 8}(2.9613,0.8830)}.&&\\
\hline
19 &$(0,0,0,0,0,0,0,\frac{1}{2},\frac{1}{2},0,0,\frac{1}{2},\frac{1}{2})$         &        &  0.222                     \\
&{\footnotesize {\bf 3}(3.1509,0.8013), {\bf 7/2}(2.7463,0.8556), {\bf 4}(3.0795,0.8286), {\bf 9/2}(2.5403,0.8985)}&&\\
&{\footnotesize {\bf 5}(2.9453,0.8993), {\bf 11/2}(2.9889,0.8381), {\bf 6}(3.1967,0.8060), {\bf 13/2}(3.0732,0.8256)}&&\\
&{\footnotesize {\bf 7}(3.1224,0.8546), {\bf 15/2}(3.0099,0.8798), {\bf 8}(2.9717,0.8935)}.&&\\
\hline
20 &$(0,0,0,\frac{1}{\sqrt{2}},\frac{1}{\sqrt{2}},0,0,0,0,0,0,0,0)$            &        &  0.221                     \\
&{\footnotesize {\bf 3}(2.6809,0.7777), {\bf 7/2}(2.6713,0.9003), {\bf 4}(3.3986,0.8664), {\bf 9/2}(2.5807,0.8817)}&&\\
&{\footnotesize {\bf 5}(3.1264,0.8965), {\bf 11/2}(3.0680,0.8894), {\bf 6}(3.1789,0.8280), {\bf 13/2}(3.1162,0.7826)}&&\\
&{\footnotesize {\bf 7}(3.1029,0.7905), {\bf 15/2}(3.0760,0.8365), {\bf 8}(3.0453,0.8659)}.&&\\
\hline

\end{tabular}
\end{center}
\end{table*}
\end{small}
\clearpage

\begin{small}
\begin{table*}[htbp]
{\Large {\bf Appendix} -- cont.}
\begin{center}
\vspace*{-0.2cm}

\begin{tabular}{llll}
\hline
No. &${\cal H}_{\rm CF}^{(6)}$      &              &  $|{{\cal H}_{\rm CF}^{(6)}}|_{\rm av}$  \\
\hline
21 &$(0,0,0,0,0,\frac{1}{\sqrt{2}},\frac{1}{\sqrt{2}},0,0,0,0,0,0)$        &        &  0.221                      \\
&{\footnotesize {\bf 3}(2.3934,0.9120), {\bf 7/2}(2.8284,0.7071), {\bf 4}(2.5041,0.8156), {\bf 9/2}(2.3351,0.8939)}&&\\
&{\footnotesize {\bf 5}(2.7121,0.7857), {\bf 11/2}(3.1849,0.7981), {\bf 6}(3.6777,0.8190), {\bf 13/2}(3.3025,0.8121)}&&\\
&{\footnotesize {\bf 7}(3.0819,0.7890), {\bf 15/2}(2.8888,0.7903), {\bf 8}(2.7651,0.8280)}.&&\\
\hline
22 &$(0,0,0,0,0,\frac{1}{2},\frac{1}{2},0,0,\frac{1}{2},\frac{1}{2},0,0)$         &        &  0.221                      \\
&{\footnotesize {\bf 3}(2.6691,0.8442), {\bf 7/2}(2.5322,0.9484), {\bf 4}(3.5241,0.8134), {\bf 9/2}(2.6315,0.8768)}&&\\
&{\footnotesize {\bf 5}(3.1247,0.9040), {\bf 11/2}(3.0538,0.8880), {\bf 6}(3.3670,0.8125), {\bf 13/2}(3.1609,0.8108)}&&\\
&{\footnotesize {\bf 7}(3.3039,0.8267), {\bf 15/2}(3.1123,0.8524), {\bf 8}(3.0357,0.8667)}.&&\\
\hline
23 &$(0,\frac{1}{\sqrt{6}},\frac{1}{\sqrt{6}},0,0,\frac{1}{\sqrt{6}},\frac{1}{\sqrt{6}},0,0,\frac{1}{\sqrt{6}},\frac{1}{\sqrt{6}},0,0)$        &        &  0.220                    \\
&{\footnotesize {\bf 3}(3.0431,0.8223), {\bf 7/2}(2.8187,0.7638), {\bf 4}(2.8556,0.8751), {\bf 9/2}(3.0500,0.8586)}&&\\
&{\footnotesize {\bf 5}(3.0159,0.8454), {\bf 11/2}(3.1400,0.8281), {\bf 6}(3.5555,0.8099), {\bf 13/2}(3.2270,0.7987)}&&\\
&{\footnotesize {\bf 7}(3.1713,0.8113), {\bf 15/2}(2.9666,0.8206), {\bf 8}(2.9375,0.8399)}.&&\\
\hline
24 &$(0,0,0,0,0,0,0,\frac{1}{\sqrt{2}},\frac{1}{\sqrt{2}},0,0,0,0)$         &        &  0.219                      \\
&{\footnotesize {\bf 3}(2.7634,0.9044), {\bf 7/2}(2.3356,0.9828), {\bf 4}(3.6181,0.7298), {\bf 9/2}(2.6630,0.8737)}&&\\
&{\footnotesize {\bf 5}(2.7589,0.8352), {\bf 11/2}(3.0346,0.7842), {\bf 6}(3.5205,0.7887), {\bf 13/2}(3.2465,0.8359)}&&\\
&{\footnotesize {\bf 7}(3.3824,0.8334), {\bf 15/2}(3.0871,0.8450), {\bf 8}(2.9640,0.8673)}.&&\\
\hline
25 &$(0,\frac{1}{\sqrt{2}},\frac{1}{\sqrt{2}},0,0,0,0,0,0,0,0,0,0)$        &   &  0.219                      \\
&{\footnotesize {\bf 3}(3.4943,0.7288), {\bf 7/2}(2.7932,0.8097), {\bf 4}(2.8761,0.8405), {\bf 9/2}(2.5996,0.8802)}&&\\
&{\footnotesize {\bf 5}(2.6356,0.8490), {\bf 11/2}(2.7690,0.8643), {\bf 6}(2.8418,0.8762), {\bf 13/2}(2.8573,0.8715)}&&\\
&{\footnotesize {\bf 7}(2.8585,0.8923), {\bf 15/2}(2.8484,0.8999), {\bf 8}(2.8246,0.8830)}.&&\\
\hline
26 &$(\frac{1}{\sqrt{5}},0,0,\frac{1}{\sqrt{5}},\frac{1}{\sqrt{5}},0,0,\frac{1}{\sqrt{5}},\frac{1}{\sqrt{5}},0,0,0,0)$         &        &  0.216                    \\
&{\footnotesize {\bf 3}(3.5153,0.7612), {\bf 7/2}(2.7902,0.8036), {\bf 4}(2.9930,0.8340), {\bf 9/2}(2.5985,0.8928)}&&\\
&{\footnotesize {\bf 5}(2.6571,0.8801), {\bf 11/2}(2.7753,0.8843), {\bf 6}(2.9523,0.8905), {\bf 13/2}(2.8492,0.8796)}&&\\
&{\footnotesize {\bf 7}(2.7998,0.8839), {\bf 15/2}(2.7503,0.8870), {\bf 8}(2.7041,0.8682)}.&&\\
\hline
27 &$(\frac{1}{\sqrt{13}},\frac{1}{\sqrt{13}},\frac{1}{\sqrt{13}},\frac{1}{\sqrt{13}},\frac{1}{\sqrt{13}},\frac{1}{\sqrt{13}},\frac{1}{\sqrt{13}},\frac{1}{\sqrt{13}},\frac{1}{\sqrt{13}},\frac{1}{\sqrt{13}},\frac{1}{\sqrt{13}},\frac{1}{\sqrt{13}},\frac{1}{\sqrt{13}})$        &        &  0.213                      \\
&{\footnotesize {\bf 3}(3.5964,0.7221), {\bf 7/2}(2.8106,0.7801), {\bf 4}(2.9216,0.8405), {\bf 9/2}(2.6110,0.8802)}&&\\
&{\footnotesize {\bf 5}(2.7647,0.9040), {\bf 11/2}(2.7428,0.9205), {\bf 6}(3.0797,0.9490), {\bf 13/2}(2.7858,0.8931)}&&\\
&{\footnotesize {\bf 7}(2.8319,0.8909), {\bf 15/2}(2.7272,0.8783), {\bf 8}(2.8052,0.8667)}.&&\\
\hline
28 &$(0,0,0,0,0,0,0,0,0,\frac{1}{\sqrt{2}},\frac{1}{\sqrt{2}},0,0)$         &        &  0.212                     \\
&{\footnotesize {\bf 3}(2.6462,0.7555), {\bf 7/2}(2.8289,0.7067), {\bf 4}(2.6836,0.8945), {\bf 9/2}(2.4491,0.8893)}&&\\
&{\footnotesize {\bf 5}(3.1151,0.8730), {\bf 11/2}(3.0875,0.7769), {\bf 6}(3.3267,0.7761), {\bf 13/2}(3.1339,0.7393)}&&\\
&{\footnotesize {\bf 7}(3.1964,0.7848), {\bf 15/2}(3.0027,0.8235), {\bf 8}(2.9881,0.8578)}.&&\\
\hline
29 &$(\frac{1}{\sqrt{2}},0,0,\frac{1}{2},\frac{1}{2},0,0,0,0,0,0,0,0)$        &        &  0.207                     \\
&{\footnotesize {\bf 3}(3.6622,0.6773), {\bf 7/2}(2.8157,0.7598), {\bf 4}(2.8102,0.8264), {\bf 9/2}(2.6680,0.9099)}&&\\
&{\footnotesize {\bf 5}(2.6464,0.9136), {\bf 11/2}(2.6129,0.9318), {\bf 6}(2.6533,0.9256), {\bf 13/2}(2.6334,0.9052)}&&\\
&{\footnotesize {\bf 7}(2.6155,0.8979), {\bf 15/2}(2.5960,0.8826), {\bf 8}(2.5837,0.8652)}.&&\\
\hline
30 &$(0,0,0,0,0,0,0,0,0,0,0,\frac{1}{\sqrt{2}},\frac{1}{\sqrt{2}})$         &        &  0.195                     \\
&{\footnotesize {\bf 3}(3.7414,0.5342), {\bf 7/2}(2.8289,0.7067), {\bf 4}(2.8978,0.8091), {\bf 9/2}(2.6452,0.8757)}&&\\
&{\footnotesize {\bf 5}(2.6918,0.9220), {\bf 11/2}(2.5667,0.9555), {\bf 6}(2.6000,0.9802), {\bf 13/2}(2.5241,0.9147)}&&\\
&{\footnotesize {\bf 7}(2.5499,0.8895), {\bf 15/2}(2.4979,0.8639), {\bf 8}(2.5183,0.8399)}.&&\\
\hline

\end{tabular}
\end{center}
\end{table*}
\end{small}
\clearpage


\renewcommand{\baselinestretch}{1}
\clearpage
\begin{small}
\begin{table*}[htbp]
\begin{center}
\caption{Extreme total splittings $\Delta E_{min}$, $\Delta E_{max}$, the averages $\overline{\Delta E}$ and the
mean square deviations $\delta$ for $|J\rangle$ states exposed to the action of 8 representative ${\cal H}_{\rm
CF}^{(2)}$s listed in Appendix, all values are given in $\sigma$. Please note, that in Tables 1 -- 3 certain
extreme values are given with the accuracy of $10^{-4}\sigma$, those found only in the representative ${\cal
H}_{\rm CF}^{(k)}$ sets with the accuracy of $10^{-2}\sigma$, but their averages and mean square deviations the
accuracy of $10^{-3}\sigma$.}

\vspace*{0.6cm}

\begin{tabular}{lcccl}
\hline
J  & $\Delta E_{min}$ & No. of ${\cal H}_{\rm CF}^{(2)}$ acc. to &  $\overline{\Delta E}$  & $\delta$ \\
 & $\Delta E_{max}$   &  the list in Appendix   &    &\\
\hline
1   & 2.1213 & 1   & 2.343  & 0.134 \\
    & 2.4495 & 8,7 &         &\\
\hline
3/2 & 2.0000 & 1   & 2.0000  & 0.0000 \\
    & 2.0000 & 8,7 &         &        \\
\hline
2   & 2.3905 & every   & 2.3905  & 0.0000 \\
    & 2.3905 &      &         &  \\
\hline
5/2 & 2.4056 & 1    & 2.433  & 0.018 \\
    & 2.4495 & 8,7  &         &  \\
\hline
3   & 2.5983 & 1    & 2.620  & 0.014 \\
    & 2.6332 & 8,7  &         &  \\
\hline
7/2 & 2.6184 & 1    & 2.693  & 0.049 \\
    & 2.7373 & 8,7  &         &  \\
\hline
4   & 2.7356 & 1    & 2.805  & 0.046 \\
    & 2.8470 & 8,7  &         &  \\
\hline
9/2 & 2.7540 & 1    & 2.867  & 0.073 \\
    & 2.9310 & 8,7  &         &  \\
\hline
5   & 2.8308 & 1    & 2.942  & 0.072 \\
    & 3.0058 & 8,7  &         &  \\
\hline
11/2& 2.8443 & 1    & 2.990  & 0.092 \\
    & 3.0682 & 8,7  &         &  \\
\hline
6   & 2.9008 & 1    & 3.045  & 0.091 \\
    & 3.1233 & 8,7  &         &  \\
\hline
13/2& 2.9124 & 1    & 3.082  & 0.106 \\
    & 3.1709 & 8,7  &         &  \\
\hline
7   & 2.9549 & 1    & 3.124  & 0.106 \\
    & 3.2138 & 8,7  &         &  \\
\hline
15/2& 2.9641 & 1    & 3.153  & 0.117 \\
    & 3.2503 & 8,7  &         &  \\
\hline
8   & 2.9973 & 1    & 3.185  & 0.117 \\
    & 3.2840 & 8,7  &         &  \\
\hline
\end{tabular}
\end{center}
\end{table*}
\end{small}

 \clearpage

\renewcommand{\baselinestretch}{1}
\clearpage
\begin{small}
\begin{table*}[htbp]
\begin{center}
\caption{Extreme total splittings $\Delta E_{min}$, $\Delta E_{max}$, the averages $\overline{\Delta E}$ and the
mean square deviations $\delta$ for $|J\rangle$ states exposed to the action of 25 representative ${\cal H}_{\rm
CF}^{(4)}$s listed in Appendix, all values are given in $\sigma$.}

\vspace*{0.2cm}

\begin{tabular}{llccl}
\hline
J  & $\Delta E_{min}$ & No. of ${\cal H}_{\rm CF}^{(4)}$ acc. to &  $\overline{\Delta E}$  & $\delta$ \\
 & $\Delta E_{max}$   &  the list in Appendix   &    &\\
\hline
2   & 2.0412 & 4   & 2.691  & 0.303 \\
    & 3.1623 & 25 &         &\\
\hline
5/2 & 2.1213 & 4   & 2.385  & 0.086 \\
    & 2.4495 & 25 &         &        \\
\hline
3   & 2.52 & 25   & 2.965  & 0.265 \\
    & 3.38 & 18   &         &  \\
\hline
7/2 & 2.33 & 24,25    & 2.640  & 0.167 \\
    & 2.8284 & 18   &         &  \\
\hline
4   & 2.36 & 24,25    & 2.654  & 0.144 \\
    & 2.82 & 18   &         &  \\
\hline
9/2 & 2.29 & 25   & 2.423  & 0.062 \\
    & 2.50 & 18   &         &  \\
\hline
5   & 2.3531 & every& 2.3531  & 0.0000 \\
    & 2.3531 &      &         &  \\
\hline
11/2& 2.38 & 4    & 2.456  & 0.050 \\
    & 2.55 & 1    &         &  \\
\hline
6   & 2.53 & 4    & 2.608  & 0.0450 \\
    & 2.69 & 1    &         &  \\
\hline
13/2& 2.45 & 4    & 2.617  & 0.096 \\
    & 2.79 & 1    &         &  \\
\hline
7   & 2.45 & 4    & 2.708  & 0.134 \\
    & 3.00 & 11  &         &  \\
\hline
15/2& 2.46 & 4    & 2.706  & 0.129 \\
    & 2.90 & 19  &         &  \\
\hline
8   & 2.52 & 4    & 2.757  & 0.140 \\
    & 2.98 & 19  &         &  \\
\hline
\end{tabular}
\end{center}
\end{table*}
\end{small}
\clearpage

\renewcommand{\baselinestretch}{1}
\clearpage
\begin{small}
\begin{table*}[htbp]
\begin{center}
\caption{Extreme total splittings $\Delta E_{min}$, $\Delta E_{max}$, the averages $\overline{\Delta E}$ and the
mean square deviations $\delta$ for $|J\rangle$ states exposed to the action of 30 representative ${\cal H}_{\rm
CF}^{(6)}$s listed in Appendix, all values are given in $\sigma$.}

\vspace*{0.2cm}

\begin{tabular}{llccl}
\hline
J  & $\Delta E_{min}$ & No. of ${\cal H}_{\rm CF}^{(4)}$ acc. to &  $\overline{\Delta E}$  & $\delta$ \\
 & $\Delta E_{max}$   &  the list in Appendix   &    &\\
\hline
3   & 2.39 & 21   & 3.004  & 0.363 \\
    & 3.7417 & 30 &         &\\
\hline
7/2 & 2.22 & 16       & 2.634  & 0.166 \\
    & 2.8284 & 30,28,21 &         &        \\
\hline
4   & 2.50 & 21   & 3.114  & 0.286 \\
    & 3.62 & 24   &         &  \\
\hline
9/2 & 2.31 & 6,21    & 2.593  & 0.119 \\
    & 3.05 & 23   &         &  \\
\hline
5   & 2.60 & 16      & 2.849  & 0.172 \\
    & 3.13 & 20,28,21   &         &  \\
\hline
11/2& 2.57 & 30   & 2.963  & 0.145 \\
    & 3.18 & 21   &         &  \\
\hline
6   & 2.60 & 30   & 3.260  & 0.292 \\
    & 3.69 & 18     &         &  \\
\hline
13/2& 2.52 & 30   & 3.061  & 0.194 \\
    & 3.30 & 21   &         &  \\
\hline
7   & 2.55 & 30    & 3.089  & 0.225 \\
    & 3.44 & 9     &         &  \\
\hline
15/2& 2.50 & 30    & 2.954  & 0.150 \\
    & 3.14 & 9     &         &  \\
\hline
8   & 2.52 & 30    & 2.905  & 0.132 \\
    & 3.08 & 15,20  &         &  \\
\hline
\end{tabular}
\end{center}
\end{table*}
\end{small}
\clearpage

\renewcommand{\baselinestretch}{1}
\clearpage
\begin{small}
\begin{table*}[htbp]
\begin{center}
\caption{Extreme total splittings of $|J\rangle$ states in virtual CF fields of Hamiltonians ${\cal H}_{\rm
CF}(q=0)=B_{20}C_{0}^{(2)}+B_{40}C_{0}^{(4)}+B_{60}C_{0}^{(6)}$ yielding constant $\sigma^{2}$ in comparison
with $\Delta E$ for the pure component $2^{k}$-poles, all values are given in $\sigma$.}

\vspace*{0.7cm}

\begin{tabular}{lccccc}
\hline Quantum     &    \multicolumn{3}{c}{The total splitting $\Delta E$}       &
                 \multicolumn{2}{c}{The range of the total splittings }                  \\
number      &    \multicolumn{3}{c}{ in the $2^{k}$-pole component}
            &    \multicolumn{2}{c}{for the $2^{k}$-pole superpositions}   \\
\hline
J& $k=2$  \hspace*{0.1cm} & $k=4$ \hspace*{0.1cm}   & $k=6$  \hspace*{0.1cm}  &
\hspace*{0.7cm} $\Delta E_{min}$ \hspace*{0.7cm}& $\Delta E_{max}$                     \\
\hline
1        &  2.1213 &  ---     &   ---    &     2.1213         &   2.1213               \\
3/2      &  2.0000 &  ---     &   ---    &     2.0000         &   2.0000               \\
2        &  2.3906 &  2.6725  &   ---    &     2.0412         &   2.7386               \\
5/2      &  2.4056 &  2.3148  &   ---    &     2.1213         &   2.4495               \\
3        &  2.5984 &  2.7717  &   3.0461 &     2.0207         &   3.2404               \\
7/2      &  2.6183 &  2.5074  &   2.4373 &     2.0000         &   2.8284               \\
4        &  2.7351 &  2.6154  &   2.8317 &     2.0802         &   3.5025               \\
9/2      &  2.7528 &  2.3651  &   2.5848 &     2.1820         &   3.1424               \\
5        &  2.8307 &  2.3531  &   2.6302 &     2.2186         &   3.3534               \\
11/2     &  2.8444 &  2.5544  &   2.8953 &     2.3075         &   3.3182               \\
6        &  2.9007 &  2.6944  &   2.9641 &     2.3350         &   3.4603               \\
13/2     &  2.9125 &  2.7883  &   2.8964 &     2.3710         &   3.4518               \\
7        &  2.9551 &  2.8478  &   2.8656 &     2.4081         &   3.6067               \\
15/2     &  2.9640 &  2.8812  &   2.9306 &     2.4293         &   3.6152               \\
8        &  2.9975 &  2.8957  &   2.9004 &     2.4532         &   3.7450               \\
\hline
\end{tabular}
\end{center}
\end{table*}
\end{small}
\clearpage

\renewcommand{\baselinestretch}{1}
\clearpage
\begin{small}
\begin{table*}[htbp]
\begin{center}
\caption{Extreme total splittings of $|J\rangle$ states in virtual CF fields of Hamiltonians ${\cal H}_{\rm
CF}(q=1)=B_{21}C_{1}^{(2)}+B_{41}C_{1}^{(4)}+B_{61}C_{1}^{(6)}$ yielding constant $\sigma^{2}$ in comparison
with $\Delta E$ for the pure component $2^{k}$-poles, all values are given in $\sigma$.}

\vspace*{0.7cm}

\begin{tabular}{lccccc}
\hline Quantum     &    \multicolumn{3}{c}{The total splitting $\Delta E$}       &
                 \multicolumn{2}{c}{The range of the total splittings }                  \\
number      &    \multicolumn{3}{c}{ in the $2^{k}$-pole component}
            &    \multicolumn{2}{c}{for the $2^{k}$-pole superpositions}   \\
\hline
J& $k=2$  \hspace*{0.1cm} & $k=4$ \hspace*{0.1cm}   & $k=6$  \hspace*{0.1cm}  &
\hspace*{0.7cm} $\Delta E_{min}$ \hspace*{0.7cm}& $\Delta E_{max}$                     \\
\hline
1        &  2.4495 &  ---     &   ---    &     2.4495         &   2.4495               \\
3/2      &  2.0000 &  ---     &   ---    &     2.0000         &   2.0000               \\
2        &  2.3906 &  3.0475  &   ---    &     2.2361         &   3.1623               \\
5/2      &  2.4495 &  2.4495  &   ---    &     2.4495         &   2.4495               \\
3        &  2.6333 &  2.7114  &   3.4943 &     2.1605         &   3.7417               \\
7/2      &  2.7373 &  2.5312  &   2.7932 &     2.0000         &   2.8284               \\
4        &  2.8470 &  2.5101  &   2.8762 &     2.2354         &   3.4166               \\
9/2      &  2.9310 &  2.3803  &   2.5996 &     2.2363         &   3.1623               \\
5        &  3.0058 &  2.3531  &   2.6356 &     2.3534         &   3.4198               \\
11/2     &  3.0682 &  2.5014  &   2.7690 &     2.4059         &   3.4636               \\
6        &  3.1233 &  2.6230  &   2.8418 &     2.4835         &   3.6040               \\
13/2     &  3.1709 &  2.7176  &   2.8573 &     2.5498         &   3.7381               \\
7        &  3.2138 &  2.8071  &   2.8585 &     2.6229         &   3.8661               \\
15/2     &  3.2504 &  2.8956  &   2.8484 &     2.6316         &   3.9880               \\
8        &  3.2840 &  2.9822  &   2.8246 &     2.6695         &   4.1043               \\
\hline
\end{tabular}
\end{center}
\end{table*}
\end{small}
\clearpage

\renewcommand{\baselinestretch}{1}
\clearpage
\begin{small}
\begin{table*}[htbp]
\begin{center}
\caption{Extreme total splittings of $|J\rangle$ states in virtual CF fields of Hamiltonians ${\cal H}_{\rm
CF}(q=2)=B_{22}C_{2}^{(2)}+B_{42}C_{4}^{(4)}+B_{62}C_{2}^{(6)}$ yielding constant $\sigma^{2}$ in comparison
with $\Delta E$ for the pure component $2^{k}$-poles, all values are given in $\sigma$.}

\vspace*{0.7cm}

\begin{tabular}{lccccc}
\hline Quantum     &    \multicolumn{3}{c}{The total splitting $\Delta E$}       &
                 \multicolumn{2}{c}{The range of the total splittings }                  \\
number      &    \multicolumn{3}{c}{ in the $2^{k}$-pole component}
            &    \multicolumn{2}{c}{for the $2^{k}$-pole superpositions}   \\
\hline
J& $k=2$  \hspace*{0.1cm} & $k=4$ \hspace*{0.1cm}   & $k=6$  \hspace*{0.1cm}  &
\hspace*{0.7cm} $\Delta E_{min}$ \hspace*{0.7cm}& $\Delta E_{max}$                     \\
\hline
1        &  2.4495 &  ---     &   ---     &     2.4495         &   2.4495               \\
3/2      &  2.0000 &  ---     &   ---     &     2.0000         &   2.0000               \\
2        &  2.3906 &  2.3906  &   ---     &     2.2362         &   3.1623               \\
5/2      &  2.4495 &  2.4495  &   ---     &     2.4495         &   2.4495               \\
3        &  2.6333 &  3.3654  &   2.6809 &     2.1605         &   3.7417               \\
7/2      &  2.7373 &  2.8222  &   2.6713 &     2.0000         &   2.8284               \\
4        &  2.8470 &  2.7999  &   3.3986 &     2.2011         &   3.9518               \\
9/2      &  2.9310 &  2.4922  &   2.5807 &     2.2361         &   3.1622               \\
5        &  3.0058 &  2.3531  &   3.1263 &     2.3461         &   3.3533               \\
11/2     &  3.0682 &  2.4141  &   3.0680 &     2.3249         &   3.4455               \\
6        &  3.1233 &  2.5657  &   3.1788 &     2.4153         &   3.5953               \\
13/2     &  3.1709 &  2.5537  &   3.1163 &     2.4521         &   3.6830               \\
7        &  3.2138 &  2.6224  &   3.1028 &     2.5064         &   3.7913               \\
15/2     &  3.2504 &  2.6016  &   3.0760 &     2.5015         &   3.8838               \\
8        &  3.2840 &  2.6260  &   3.0454 &     2.5249         &   3.9745               \\
\hline
\end{tabular}
\end{center}
\end{table*}
\end{small}
\clearpage

\renewcommand{\baselinestretch}{1}
\clearpage
\begin{small}
\begin{table*}[htbp]
\begin{center}
\caption{Extreme total splittings of $|J\rangle$ states in virtual CF fields of Hamiltonians ${\cal H}_{\rm
CF}(q=3)=B_{43}C_{3}^{(4)}+B_{63}C_{3}^{(6)}$ yielding constant $\sigma^{2}$ in comparison with $\Delta E$ for
the pure component $2^{k}$-poles, all values are given in $\sigma$.}

\vspace*{0.7cm}

\begin{tabular}{lcccc}
\hline Quantum     &    \multicolumn{2}{c}{The total splitting $\Delta E$}       &
                 \multicolumn{2}{c}{The range of the total splittings }                  \\
number      &    \multicolumn{2}{c}{ in the $2^{k}$-pole component}
            &    \multicolumn{2}{c}{for the $2^{k}$-pole superpositions}   \\
\hline
J & $k=4$ \hspace*{0.1cm}   & $k=6$  \hspace*{0.1cm}  &
\hspace*{0.7cm} $\Delta E_{min}$ \hspace*{0.7cm}& $\Delta E_{max}$                     \\
\hline
2        &  2.2361 &  ---     &     2.2361         &   2.2361               \\
5/2      &  2.4495 &  ---     &     2.4495         &   2.4495               \\
3        &  3.3845 &  2.3934  &     2.1605         &   3.7417               \\
7/2      &  2.8284 &  2.8284  &     2.8284         &   2.8284               \\
4        &  2.8161 &  2.5041  &     2.4499         &   3.6586               \\
9/2      &  2.4950 &  2.3351  &     2.2362         &   3.1622               \\
5        &  2.3531 &  2.7121  &     2.3469         &   3.5904               \\
11/2     &  2.4037 &  3.1850  &     2.3301         &   3.2130               \\
6        &  2.5343 &  3.6777  &     2.4945         &   3.7054               \\
13/2     &  2.5279 &  3.3025  &     2.4540         &   3.3101               \\
7        &  2.6131 &  3.0819  &     2.4537         &   3.0833               \\
15/2     &  2.5620 &  2.8888  &     2.4600         &   2.9024               \\
8        &  2.5827 &  2.7651  &     2.5079         &   2.9333               \\
\hline
\end{tabular}
\end{center}
\end{table*}
\end{small}
\clearpage

\renewcommand{\baselinestretch}{1}
\clearpage
\begin{small}
\begin{table*}[htbp]
\begin{center}
\caption{Extreme total splittings of $|J\rangle$ states in virtual CF fields of Hamiltonians ${\cal H}_{\rm
CF}(q=4)=B_{44}C_{4}^{(4)}+B_{64}C_{4}^{(6)}$ yielding constant $\sigma^{2}$ in comparison with $\Delta E$ for
the pure component $2^{k}$-poles, all values are given in $\sigma$.}

\vspace*{0.7cm}

\begin{tabular}{lcccc}
\hline Quantum     &    \multicolumn{2}{c}{The total splitting $\Delta E$}       &
                 \multicolumn{2}{c}{The range of the total splittings }                  \\
number      &    \multicolumn{2}{c}{ in the $2^{k}$-pole component}
            &    \multicolumn{2}{c}{for the $2^{k}$-pole superpositions}   \\
\hline
J & $k=4$ \hspace*{0.1cm}   & $k=6$  \hspace*{0.1cm}  &
\hspace*{0.7cm} $\Delta E_{min}$ \hspace*{0.7cm}& $\Delta E_{max}$                     \\
\hline
2        &  3.1623 &  ---     &     3.1623         &   3.1623               \\
5/2      &  2.4495 &  ---     &     2.4495         &   2.4495               \\
3        &  2.5225 &  2.7634  &     2.1605         &   3.7417               \\
7/2      &  2.3352 &  2.3355  &     2.0000         &   2.8284               \\
4        &  2.3796 &  3.6181  &     2.1857         &   4.0763               \\
9/2      &  2.2901 &  2.6629  &     2.2362         &   3.1623               \\
5        &  2.3531 &  2.7589  &     2.3514         &   3.0557               \\
11/2     &  2.4838 &  3.0346  &     2.4495         &   3.1697               \\
6        &  2.5960 &  3.5206  &     2.3785         &   3.6621               \\
13/2     &  2.6614 &  3.2465  &     2.4176         &   3.3513               \\
7        &  2.7386 &  3.3824  &     2.4212         &   3.4501               \\
15/2     &  2.8020 &  3.0871  &     2.3992         &   3.1517               \\
8        &  2.8684 &  2.9640  &     2.3673         &   3.0352               \\
\hline
\end{tabular}
\end{center}
\end{table*}
\end{small}
\clearpage

\renewcommand{\baselinestretch}{1}
\clearpage
\begin{small}
\begin{table*}[htbp]
\begin{center}
\caption{Total splitting $\Delta E$ (in $\sigma$) of $|J\rangle$ states in virtual CF fields of partial
Hamiltonians ${\cal H}_{\rm CF}^{(6)}=B_{65}C_{5}^{(6)}$ and ${\cal H}_{\rm CF}^{(6)}=B_{66}C_{6}^{(6)}$
yielding the same $\sigma^{2}$.}

\vspace*{0.7cm}

\begin{tabular}{lcc}
\hline Quantum     &    \multicolumn{2}{c}{$\Delta E$}               \\
                number   &                   \\
J      &    $q=5$
            &    $q=6$   \\
\hline
3        &  2.6458 &  3.7417                 \\
7/2      &  2.8284 &  2.8284                 \\
4        &  2.6835 &  2.8980                 \\
9/2      &  2.4495 &  2.6454                 \\
5        &  3.1152 &  2.6919                          \\
11/2     &  3.0872 &  2.5667                          \\
6        &  3.3269 &  2.6002                          \\
13/2     &  3.1339 &  2.5241                        \\
7        &  3.1964 &  2.5499              \\
15/2     &  3.0030 &  2.4979                \\
8        &  2.9881 &  2.5183                     \\
\hline
\end{tabular}
\end{center}
\end{table*}
\end{small}
\clearpage

\renewcommand{\baselinestretch}{1.2}
\clearpage
\begin{small}
\begin{table*}[htbp]
\begin{center}
\caption{Nominally extreme total splittings $\Delta {\cal E}$ (conserving $\sigma^{2}$) of states $|J\rangle$
expressed in $\sigma$ .}

\vspace*{0.7cm}

\begin{tabular}{lrr}
\hline
Quantum number J      &    $\Delta {\cal E}_{min}$            &   $\Delta {\cal E}_{max}$   \\
\hline
1        &  $3/\sqrt{2}=2.1213$  &  $\sqrt{6}=2.4495$                        \\
3/2      &  2.0000 &  2.0000                                                 \\
2        &  $5/\sqrt{6}=2.0412$  &  $\sqrt{10}=3.1623$                       \\
5/2      &  $3/\sqrt{2}=2.1213$  &  $\sqrt{6}=2.4495$                        \\
3        &  $7/\sqrt{12}=2.0207$ &  $\sqrt{14}=3.7417$                       \\
7/2      &  2.0000               &  $\sqrt{8}=2.8284 $                                                \\
4        &  $9/\sqrt{20}=2.0125$ &  $\sqrt{18}=4.2426$                       \\
9/2      &  $5/\sqrt{6}=2.0412$  &  $\sqrt{10}=3.1623$                                                 \\
5        &  $11/\sqrt{30}=2.0083$&  $\sqrt{22}=4.6904$                        \\
11/2     &  2.0000               &  $\sqrt{12}=3.4641$                                                  \\
6        &  $13/\sqrt{42}=2.0059$&  $\sqrt{26}=5.0990$                                              \\
13/2     &  $7/\sqrt{12}=2.0207$ &  $\sqrt{14}=3.7417$                                                    \\
7        &  $15/\sqrt{56}=2.0045$&  $\sqrt{30}=5.4772$                                                \\
15/2     &  $2.0000$             &  $\sqrt{16}=4.0000$                                                    \\
8        &  $17/\sqrt{72}=2.0035$&  $\sqrt{34}=5.8310$                                                  \\
\hline
\end{tabular}
\end{center}
\end{table*}
\end{small}
\clearpage

\renewcommand{\baselinestretch}{1.2}
\clearpage
\begin{small}
\begin{table*}[htbp]
\begin{center}
\caption{Numerically estimated extreme total splittings $\Delta E_{min}$ and $\Delta E_{max}$ of $|J\rangle$
states in crystal-fields given in relation to $\sigma$.}

\vspace*{0.7cm}

\begin{tabular}{lrr}
\hline
Quantum number J      &    $\Delta E_{min}$            &  $\Delta E_{max}$   \\
\hline
1        &  2.1213  &  2.4495                      \\
3/2      &  2.0000  &  2.0000                      \\
2        &  2.0412  &  3.1623                      \\
5/2      &  2.1213  &  2.4495                      \\
3        &  2.0207  &  3.7417                      \\
7/2      &  2.0000  & 2.8284                       \\
4        &  2.0802  &  4.0763                      \\
9/2      &  2.1820  &  3.1623                      \\
5        &  2.2186  &  3.5904                      \\
11/2     &  2.3075  &  3.4636                      \\
6        &  2.3350  &  3.7054                      \\
13/2     &  2.3710  &  3.7381                      \\
7        &  2.4081  &  3.8661                      \\
15/2     &  2.3993  &  3.9880                      \\
8        &  2.3673  &  4.1043                      \\
\hline
\end{tabular}
\end{center}
\end{table*}
\end{small}
\clearpage

\end{document}